\documentclass[10pt]{article}
\usepackage{amsmath}
\usepackage{latexsym}
\usepackage{amssymb}
\usepackage{amsfonts}
\usepackage{amsthm}
\title{KILLING OPERATOR FOR THE KERR METRIC}
\author{J.-F. Pommaret \\ CERMICS, Ecole des Ponts ParisTech, Paris, France \\
 jean-francois.pommaret@wanadoo.fr,  \\
ORCID: 0000-0003-0907-2601}
\date{  }
\begin{document}
\maketitle

\noindent
{\bf ABSTRACT} \\

\noindent
When ${\cal{D}}: E \rightarrow F$ is a linear differential operator of order $q$  between the sections of vector bundles over a manifold $X$ of dimension $n$, it is defined by a bundle map $\Phi: J_q(E) \rightarrow F=F_0$ that may depend, explicitly or implicitly, on constant parameters $a, b, c, ...$. A "direct problem " is to find the generating compatibility conditions (CC) in the form of an operator ${\cal{D}}_1: F_0 \rightarrow F_1$. When ${\cal{D}}$ is involutive, that is when the corresponding system $R_q=ker(\Phi)$ is involutive, this procedure provides successive first order involutive operators ${\cal{D}}_1, ... , {\cal{D}}_n$ . Though ${\cal{D}}_1 \circ {\cal{D}}=0 $ implies $ad({\cal{D}}) \circ ad({\cal{D}}_1)=0$ by taking the respective adjoint operators, then $ad({\cal{D}})$ may not generate the CC of $ad({\cal{D}}_1)$ and measuring such "gaps" led to introduce extension modules in differential homological algebra. They may also depend on the parameters and such a situation is well known in ordinary or partial control theory. When $R_q$ is not involutive, a standard {\it prolongation/projection} (PP) procedure allows in general to find integers $r,s$ such that the image $R^{(s)}_{q+r}$ of the projection at order $q+r$ of the prolongation ${\rho}_{r+s}(R_q) = J_{r+s}(R_q) \cap J_{q+r+s}(E)\subset J_{r+s}(J_q(E)) $ is involutive but it may highly depend on the parameters. However, sometimes the resulting system no longer depends on the parameters and the extension modules do not depend on the parameters because it is known that they do not depend on the differential sequence used for their definition. The purpose of this paper is to study the above problems for the Kerr $(m, a)$, Schwarzschild $(m, 0)$ and Minkowski $(0, 0)$ parameters while computing the dimensions of the inclusions 
$R^{(3)}_1\subset R^{(2)}_1 \subset R^{(1)}_1 =R_1 \subset J_1(T(X))$ for the respective Killing operators. Other striking motivating examples are also presented.  \\ \\

\newpage

\noindent
{\bf 1) INTRODUCTION}. \\

With standard notations of differential geometry, let $(E,F, ...)$ be vector bundles over a manifold $X$ with dimension $n$ with sections $(\xi, \eta, ... )$, tangent bundle $T$ and cotangent bundle $T^*$. We shall denote by $J_q(E) $ the $q$-jet bundle of $E$ with sections ${\xi}_q$ transforming like the $q$-derivatives $j_q(\xi)$. If $\Phi: J_q(E) \rightarrow F$ is a bundle morphism, we shall consider the system $R_q = ker(\Phi) \subset J_q(E)$ of order $q$ on $E$. The $r$-prolongation ${\rho}_r(R_q)= J_r(R_q) \cap J_{q+r}(E) \subset  J_r(J_q(E))$ obtained by differentiating formally $r$ times the given  ordinary (OD) or partial (PD) defining equations of $R_q$ will be the kernel of the composite morphism 
${\rho}_r(\Phi): J_{q++r}(E) \rightarrow J_r(J_q(E)) \rightarrow J_r(F)$. The symbol $g_{q+r} = R_{q+r} \cap S_{q+r}T^* \otimes E  \subset J_{q+r}(E) $ of $R_{q+r}$ is the $r$-prolongation of the symbol $g_q$ of $R_q$ and the kernel of the composite morphism ${\sigma}_r(\Phi): S_{q+r}T^* \otimes E \rightarrow S_rT^* \otimes F$ obtained by restriction. The Spencer operator $d: R_{q+1} \rightarrow T^* \otimes R_q: {\xi}_{q+1} \rightarrow j_1 ({\xi}_q) - {\xi}_{q+1}$ is obtained by using the fact that $R_{q+1} = J_1(R_q) \cap J_{q+1}(E)$ and that $J_1(R_q)$ is an affine vector bundle over $R_q$ modelled on $T^* \otimes R_q$. We shall always suppose that $\Phi$ is an epimorphism and introduce the vector bundle 
$F_0 = J_q(E)/ R_q$. The system $R_q$ is aid to be {\it formally integrable} (FI) if $r+1$ prolongations do not bring new equations of order $q+r$ other than the ones obtained after only $r$ prolongations, for any $r\geq 0$, that is all the equations of order $q+r$ can be obtained by differentiating $r$ times {\it only} the given equations of order $q$ for any $r \geq 0$. The system is said to be " {\it involutive} " if it is FI and the symbol $g_q$ is involutive, a purely algebraic property [7, 9]. In that case, the 
successive CC operators can only be {\it at most} ${\cal{D}}_1, ... , {\cal{D}}_n$ which are first order and involutive operators. \\

The next problem is to define the CC operator ${\cal{D}}_1: F_0 \rightarrow F_1: \eta \rightarrow \zeta$ in such a way that the CC of ${\cal{D}} \xi = \eta$ is of the form 
${\cal{D}}_1 \eta = 0$. As shown in many books [7-11, 14, 15] and papers [16, 18, 22], such a problem may be quite difficult because the order of the generating CC may be quite high. Proceeding in this way, we may construct the CC ${\cal{D}}_2: F_1 \rightarrow F_2$ of ${\cal{D}}_1$ and so on. A second problem shown on the motivating examples is that " {\it jumps} " in the successive orders may appear, even on elementary examples. Now, if the map $\Phi$ depends on constant (or variable) parameters $(a, b, c, ... )$, then the study of the two previous problems becomes much harder because the ranks of the matrices ${\rho}_r(\Phi)$ and/or ${\sigma}_r(\Phi)$ may also highly depend on the parameters as we shall see. The case of the Killing operator for the Minkowski metric is well known  but the study of the Killing operator for the Schwarzschild and Kerr metrics is rather recent and striking as it proves that both the numbers of generating second order CC and the numbers of generating third order CC may change [1-4]. \\

The purpose of the present paper is to revisit these works by using new homological techniques [10, 15, 20]. It is a matter of fact that they do not agree with the previous ones for the third order CC. In order to escape from such an unpleasant situation, we have written this paper in such a way that we are only using elementary combinatorics and diagram chasing. However, an equally important second purpose is to question the proper target of the quoted publications. Indeed, important concepts such as {\it differential extension modules} have been introduced in {\it differential homological algebra} and are known, thanks to a quite difficult theorem [10, 25], to be the only intrinsic results that could be obtained independently of the differential sequence that could be used, provided that $\Theta= \ker({\cal{D}})$ is the same, that is even if one is using another system on $E$ with the same solutions. Equivalently, this amounts to say, in a few words but a more advanced language, if {\it we are keeping the same differential module} $ M $ {\it but changing its presentation}. \\

However, as we shall see, there are even simple academic systems depending on parameters but such that a convenient system (say involutive) may no longer depend on the parameters. Also, in a totally independent way still not acknowledged, E. Vessiot has shown that certain operators may depend on geometric objects satisfying non-linear {\it structure equations} that are depending on certain {\it Vessiot structure constants} $c$. The simplest example is the condition of constant Riemannian curvature [7, 24] which is necessary in order that the Killing system becomes FI but we shall prove that the case of contact structures is similar. In such situations, we shall prove that the extension modules only depend on these constants. \\

\noindent 
We now recall the main results and definitions that are absolutely needed for the applications.\\
With canonical epimorphism ${\Phi}_0=\Phi:J_q(E) \Rightarrow J_q(E)/R_q=F=F_0$, the various prolongations are described by the following commutative and 
exact "{\it introductory diagram} " used in the sequel:  \\
\[   \begin{array}{rcccccccl}
    &  0  &  &  0  &  &  0  &  &   &   \\
     &  \downarrow &  & \downarrow &  &  \downarrow  &  &   &     \\
0 \rightarrow  &  g_{q+r+1}  & \rightarrow & S_{q+r+1}T^*\otimes E & \stackrel{{\sigma}_{r+1}(\Phi)}{\longrightarrow} &  S_{r+1}T^*\otimes F_0 & \rightarrow &   h_{r+1} &  \rightarrow 0 \\
&  \downarrow &  & \downarrow &  &  \downarrow  &  &  \downarrow    &     \\
0 \rightarrow &  R_{q+r+1} & \rightarrow  & J_{q+r+1}(E) & \stackrel{{\rho}_{r+1}(\Phi)}{\longrightarrow}  &  J_{r+1}(F_0) &  \rightarrow &  
Q_{r+1}  &  \rightarrow 0  \\
    &  \downarrow &  & \downarrow &  &  \downarrow  &  & \downarrow   &     \\
0 \rightarrow &  R_{q+r} & \rightarrow  & J_{q+r}(E) & \stackrel{{\rho}_r(\Phi)}{\longrightarrow}  &  J_r(F_0) &  \rightarrow &  Q_r  & 
\rightarrow 0 \\ 
   &  &  & \downarrow &  &  \downarrow  &  & \downarrow  & \\
   &  &  &  0  && 0  && 0  &
\end{array}   \]
Chasing along the diagonal of this diagram while applying the standard "{\it snake}" lemma, we obtain the useful "{\it long exact connecting sequence} " also often used in the sequel:  \\
 \[  0  \rightarrow  g_{q+r+1}  \rightarrow R_{q+r+1}  \rightarrow R_{q+r}  \rightarrow h_{r+1} \rightarrow Q_{r+1}  \rightarrow Q_r \rightarrow 0  \]
which is thus connecting in a tricky way FI ({\it lower left}) with CC ({\it upper right}).  \\
\noindent
A key step in the procedure for constructing differential sequences will be to use the following (difficult)  theorems and corollary (For Spencer cohomology and acyclicity or involutivity  (See [7, 11] for details and compare to [5, 26]) (See [7, 9, 15, 22] for more details):  \\

\noindent
{\bf THEOREM 1.1}: There is a finite {\it Prolongation/Projection} (PP) algorithm providing two integers $r,s\geq 0$ by successive increase of each of them such that the new system $R^{(s)}_{q+r}= {\pi}^{q+r+s}_{q+r}(R_{q+r+s})$ has the same solutions as $R_q$ but is FI with a $2$-acyclic or involutive symbol and first order CC. The maximum order of ${\cal{D}}_1$ is thus equal to $ r+s+1$ as we used $r+s$ prolongations but it may be lower because certain CC may generate the higher order ones as will be seen in the motivating examples.   \\ 

\noindent 
{\bf DEFINITION 1.2}: A differential sequence is said to be {\it formally exact} if it is exact on the jet level composition of the prolongations involved. A formally exact sequence is 
said to be {\it strictly exact} if all the operators/systems involved are FI (See [7, 9, 13] for more details). A strictly exact sequence is called {\it canonical} if all the operators/systems 
are involutive. \\

When $d: J_{q+1}(E) \rightarrow T^* \otimes J_q(E): {\xi}_{q+1} \rightarrow j_1({\xi}_q)  - {\xi}_{q+1}$ is the Spencer operator, we have:  \\

\noindent
{\bf PROPOSITION 1.3}: If $R_q\subset J_q(E)$ and $R_{q+1} \subset J_{q+1}(E)$ are two systems of respective orders $q$ and $q+1$, then 
$R_{q+1} \subset {\rho}_1(R_q)$ if and onlty if ${\pi}^{q+1}_q (R_{q+1}) \subset R_q$ {\it and} $dR_{q+1} \subset T^* \otimes R_q$.  \\

\noindent
{\bf DEFINITION 1.4}: Let us "cut" the preceding introductory diagram  by means of a central vertical line and define $R'_r = im({\rho}_r(\Phi)) \subseteq J_r(F_0)$ with $R'_0=F_0$. Chasing in this diagram, we notice that ${\pi}^{r+1}_r:J_{r+1}(F_0) \rightarrow J_r(F_0)$ induces an epimorphism ${\pi}^{r+1}_r: R'_{r+1} \rightarrow R'_r, \forall r\geq 0$. However, a chase in this diagram proves that {\it the kernel of this epimorphism is not} $im({\sigma}_{r+1}(\Phi))$ unless $R_q$ is FI ({\it care}). For this reason, we shall define it to be {\it exactly} 
$g'_{r+1}$.  \\

\noindent
{\bf THEOREM 1.5}:  $R'_{r+1} \subseteq {\rho}_1(R'_r)$ and $ dim ({\rho}_1(R'_r)) - dim (R'_{r+1})$ is the number of new generating CC of order $r+1$ .  \\

\noindent
{\bf COROLLARY 1.6}: The system $R'_r \subset J_r(F_0)$ becomes FI with a $2$-acyclic or involutive symbol and $R'_{r+1}={\rho}_1(R'_r)  \subset J_{r+1}(F_0)$ when $r$ is large enough.  \\ \\

\noindent
{\bf  2) MOTIVATING EXAMPLES} \\

\noindent
{\bf EXAMPLE 2.1}: Let $n=2, m= 1$ and introduce the trivial vector bundle $E$ with local coordinates $(x^1, x^2, \xi)$ for a section over the base manifold $X$ with local coordinates $(x^1,x^2)$. Let us consider the linear second order system $R_2 \subset J_2(E)$ defined by the two linearly independent equations $ d_{22}\xi= 0, \,\, d_{12}\xi + a d_1 \xi = 0$ where $a$ is an arbitrary constant parameter. Using crossed derivatives, we get the second order system $R^{(1)}_2 \subset R_2$ defined by the PD equations 
$d_{22} \xi = 0, d_{12} \xi + a d_1 \xi= 0, a^2 d_1\xi=0$ 
which is easily seen not to be involutive. Hence we have two possibilities:  \\
 $\bullet  \,\,\, a=0 $: We obtain the following second order homogeneous involutive system: \\
\[ R^{(1)}_2 = R_2 \subset J_2(E) \hspace{2cm} \left\{  \begin{array}{rcl}
 d_{22} \xi &=  & {\eta}^2 \\
d_{12} \xi  & = & {\eta}^1  
\end{array}
\right.  \fbox{ $ \begin{array}{ll}
1 & 2   \\
1 & \bullet 
\end{array} $ } \] 
with the only first order homogeneous involutive CC $d_1 {\eta}^2 - d_2 {\eta}^1= 0 $ leading to the Janet sequence:  \\
\[   0 \rightarrow  \Theta \rightarrow E \underset 2 {\stackrel{{\cal{D}}}{\longrightarrow }} F_0   \underset 1 {\stackrel{{\cal{D}}_1}{\longrightarrow }} F_1  \rightarrow 0  \]
We let the reader check as an easy exercise that $ad({\cal{D}})$ which is of order $2$ does not generate the CC of $ad({\cal{D}}_1)$ which is of order $1$. \\

\noindent
$\bullet  \,\,\, a \neq 0$: We obtain the second order system $R^{(1)}_2 $ defined by $ d_{22} \xi = 0, d_{12} \xi = 0, d_1\xi = 0 $ with a strict inclusion $R^{(1)}_2 \subset R_2$ because $3<4$. We may define $\eta = d_1{\eta}^2 - d_2 {\eta}^1+ a {\eta}^1$ and obtain the involutive and finite type system in $\delta$-regular coordinates:  \\
\noindent
\[ R^{(2)}_2 \subset J_2(E) \hspace{2cm} \left\{  \begin{array}{rcl}
 d_{22} \xi &=  & {\eta}^2  \\
 d_{12} \xi & =  & {\eta}^1 - \frac{1}{a} \eta   \\
 d_{11} \xi & =  & \frac{1}{a^2} d_1 \eta \\
 d_1  \xi    & = & \frac{1}{a^2} \eta
\end{array}
\right.  \fbox{ $ \begin{array}{ll}
1 & 2   \\
1 & \bullet \\  
1 & \bullet  \\
\bullet & \bullet 
\end{array} $ } \]
Counting the dimensions, we have the following strict inclusions by comparing the dimensions:  \\
\[      R^{(2)}_2 \subset R^{1)}_2 \subset R_2 \subset J_2(E) , \hspace{3cm}  2 <  3   <  4  <  6  \]

\noindent
{\bf LEMMA 2.1.1}: The symbol $g_{2+r} $ is involutive with $dim(g_{2+r})=1, \forall r\geq 0$. Moreover, we have $dim(R_{2+r})  = 4, \forall r\geq 0$. \\

\noindent
{\it Proof}: Using jet notation, the $4$ parametric jets of $R_2$ are $(\xi, {\xi}_1, {\xi}_2, {\xi}_{11})$. The $4$ parametric jets of $R_3$ are now $(\xi, {\xi}_2, {\xi}_{11}, {\xi}_{111})$ and so on. Accordingly, the dimension of $g_{2+r}$ is $1$ because the only parametric jet is ${\xi}_{1....1}$. We have the short exact sequence 
$0 \rightarrow g_{r+2} \rightarrow R_{r+2} \rightarrow R^{(1)}_{r+1} \rightarrow 0 $ and the symbol of $R_2$ is involutive. It follows from a delicate but crucial theorem (See [7], Theorem 2.4.5 p 70 and proposition 2.5.1 p 76 with $q=2$) that ${\rho}_r(R^{(1)}_2) = R^{(1)}_{2+r} \,\,, \forall r\geq 0$ with $dim(R^{(1)}_{r+1})= 3$, a result leading to $dim(R_{r+2})= 3+1=4$ by counting the dimensions. As $R^{(1)}_2$ does not depend any longer on the parameter, the general solution is easily seen to be of the form $\xi = cx^2 + d$ and is thus only depending on two arbitrary constants, contrary to what could be imagined from this lemma but in a coherent way with the fact that $dim(R^{(2)}_{2+r}) = 2, \forall r\geq 0$.    \\
\hspace*{12cm}    $  \Box $ \\

\noindent
After differentiating twice, we could be waiting for CC of order $3$. However, we obtain the $4$ CC:  \\
\[ d_2 {\eta}^1 - \frac{1}{a} d_2 \eta - d_1 {\eta}^2 = 0, \frac{1}{a^2} d_{12} \eta - d_1 {\eta}^1 + \frac{1}{a} d_1\eta = 0, \frac{1}{a^2} d_2 \eta - {\eta}^1 + \frac{1}{a} \eta =0, 
\frac{1}{a^2} (d_1 \eta - d_1 \eta) = 0  \]
The last CC that we shall call "{\it identity to zero} " must not be taking into account. The second CC is just the derivative with respect to $x^1 $ of the third CC which amounts to 
\[ (d_{12} {\eta}^2 - d_{22}{\eta}^1 + a d_2 {\eta}^1) - a^2 {\eta}^1 + a (d_1 {\eta}^2 - d_2 {\eta}^1 + a {\eta}^1)=0 \Leftrightarrow  d_{12} {\eta}^2 - d_{22} {\eta}^1 + a d_1 {\eta}^2=0 \]
which is a second order CC amounting to the first. Hence we get the only generating CC operator ${\cal{D}}_1: ({\eta}^1, {\eta}^2) \rightarrow d_{12} {\eta}^2 - d_{22} {\eta}^1 + a d_1 {\eta}^2=\zeta $ which is thus formally surjective.  \\

For helping the reader, we recall that basic elementary combinatorics arguments are giving $dim(S_qT^*) = q+1$ while $dim(J_q(E)) = (q+1)(q+2)/2$ because $n=2$ and 
$m = dim(E)=1$.\\

\[   \begin{array}{rcccccccl}
    &  0  &  &  0  &  &  0  &  &   &   \\
     &  \downarrow &  & \downarrow &  &  \downarrow  &  &   &     \\
0 \rightarrow  &  g_2  & \rightarrow & S_2T^* \otimes E & \longrightarrow & F_0 & \rightarrow & 0 &   \\
&  \downarrow &  & \downarrow &  &  \parallel &  &  &     \\
0 \rightarrow &  R_2 & \rightarrow  & J_2(E) & \longrightarrow &  F_0 &  \rightarrow &  0 &   \\
    &  \downarrow &  & \downarrow &  &  \downarrow  &  &   &     \\
0 \rightarrow &  R_1 & \rightarrow  & J_1 (E) & \longrightarrow  & 0 &  &  &               \\ 
   & \downarrow  &  & \downarrow &  &  \downarrow  &  &   & \\
   & 0 &  &  0  && 0  && &
\end{array}   \]

\[   \begin{array}{rcccccccl}
    &  0  &  &  0  &  &  0  &  &   &   \\
     &  \downarrow &  & \downarrow &  &  \downarrow  &  &   &     \\
0 \rightarrow  &  1 &\longrightarrow &3 & \longrightarrow &  2 & \rightarrow &   0 &  \\
&  \downarrow &  & \downarrow &  &  \parallel  &  &     &     \\
0 \rightarrow &  4 & \rightarrow  & 6 & \longrightarrow & 2  &  \rightarrow & 0 &  \\
    &  \downarrow &  & \downarrow &  &  \downarrow  &  &   &     \\
0 \rightarrow & 3 & \rightarrow  & 3 & \longrightarrow  &  0 &   &   & \\ 
   & \downarrow  &  & \downarrow &  &  \downarrow  &  &  & \\
   & 0 &  &  0  && 0  &&  &
\end{array}   \]

\[   \begin{array}{rcccccccl}
    &  0  &  &  0  &  &  0  &  &   &   \\
     &  \downarrow &  & \downarrow &  &  \downarrow  &  &   &     \\
0 \rightarrow  &  g_3  & \rightarrow & S_3T^* \otimes E & \rightarrow &  T^*\otimes F_0 & \rightarrow &   h_1 &  \rightarrow 0 \\
&  \downarrow &  & \downarrow &  &  \downarrow  &  &  \downarrow    &     \\
0 \rightarrow &  R_3 & \rightarrow  & J_3(E) & \rightarrow &  J_1(F_0) &  \rightarrow &  Q_1 & \rightarrow 0   \\
    &  \downarrow &  & \downarrow &  &  \downarrow  &  & \downarrow   &     \\
0 \rightarrow &  R_2 & \rightarrow  & J_2 (E) & \rightarrow &  F_0 &  \rightarrow &  0 &               \\ 
   &  &  & \downarrow &  &  \downarrow  &  &   & \\
   &  &  &  0  && 0  && &
\end{array}   \]

\[   \begin{array}{rcccccccl}
    &  0  &  &  0  &  &  0  &  &   &   \\
     &  \downarrow &  & \downarrow &  &  \downarrow  &  &   &     \\
0 \rightarrow  &  1 &\longrightarrow &4 & \rightarrow &  4 & \rightarrow &   1 &  \rightarrow 0 \\
&  \downarrow &  & \downarrow &  &  \downarrow  &  &  \downarrow    &     \\
0 \rightarrow &  4 & \rightarrow  & 10 & \rightarrow & 6  &  \rightarrow & 0 &  \\
    &  \downarrow &  & \downarrow &  &  \downarrow  &  &   &     \\
0 \rightarrow & 4 & \rightarrow  & 6 & \rightarrow  &  2 &  \rightarrow &  0 & \\ 
   &  &  & \downarrow &  &  \downarrow  &  &  & \\
   &  &  &  0  && 0  &&  &
\end{array}   \]

\[   \begin{array}{rcccccccl}
    &  0  &  &  0  &  &  0  &  &   &   \\
     &  \downarrow &  & \downarrow &  &  \downarrow  &  &   &     \\
0 \rightarrow  &  g_4  & \rightarrow & S_4T^* \otimes E & \rightarrow &  S_2 T^*\otimes F_0 & \rightarrow &   h_2 &  \rightarrow 0 \\
&  \downarrow &  & \downarrow &  &  \downarrow  &  &  \downarrow    &     \\
0 \rightarrow &  R_4 & \rightarrow  & J_4(E) & \rightarrow &  J_2(F_0) &  \rightarrow &  Q_2 & \rightarrow 0   \\
    &  \downarrow &  & \downarrow &  &  \downarrow  &  & \downarrow   &     \\
0 \rightarrow &  R_3 & \rightarrow  & J_3(E) & \rightarrow &  J_1(F_0) &  \rightarrow &  0 &               \\ 
   &  &  & \downarrow &  &  \downarrow  &  &   & \\
   &  &  &  0  && 0  && &
\end{array}   \]

\[   \begin{array}{rcccccccl}
    &  0  &  &  0  &  &  0  &  &   &   \\
     &  \downarrow &  & \downarrow &  &  \downarrow  &  &   &     \\
0 \rightarrow  &  1 & \rightarrow &5 & \rightarrow &  6 & \rightarrow &   2 &  \rightarrow 0 \\
&  \downarrow &  & \downarrow &  &  \downarrow  &  &  \downarrow    &     \\
0 \rightarrow &  4 & \rightarrow  & 15 & \rightarrow & 12  &  \rightarrow & 1 &  \rightarrow 0  \\
    &  \downarrow &  & \downarrow &  &  \downarrow  &  & \downarrow   &     \\
0 \rightarrow & 4 & \rightarrow  & 10 & \rightarrow  &  6 &  \rightarrow &  0 & \\ 
   &  &  & \downarrow &  &  \downarrow  &  &  & \\
   &  &  &  0  && 0  &&  &
\end{array}   \]
Using these diagrams, we obtain successively till we stop:
\[ R'_0=F_0  \Rightarrow R'_1=J_1(F_0) \Rightarrow  {\rho}_1(R'_1) = J_2(F_0) \Rightarrow R'_2 \subset {\rho}_1(R'_1) \Rightarrow  R'_3={\rho}_1(R'_2) \]
Hence, the number of generating CC of order $1$ is zero and the number of generating CC of strict order $2$ is $dim({\rho}_1(R'_1)) - dim(R'_2) = 12 - (15-4)=12-11=1$ in a 
coherent way.    \\
Setting $F_1 = Q_2$ with $dim(Q_2) = 1$, we obtain the commutative diagram:  \\

\[   \begin{array}{rcccccccl}
    &  0  &  &  0  &  &  0  &  &  0  &   \\
     &  \downarrow &  & \downarrow &  &  \downarrow  &  & \downarrow   &     \\
0 \rightarrow  &  g_5  & \rightarrow & S_5T^* \otimes E & \rightarrow &  \fbox { $ S_3 T^*\otimes F_0 $} & \rightarrow &  T^* \otimes F_1 & \rightarrow 0  \\
&  \downarrow &  & \downarrow &  &  \downarrow  &  &  \downarrow    &     \\
0 \rightarrow &  R_5 & \rightarrow  & J_5(E) & \rightarrow &  J_3(F_0) &  \rightarrow &  J_1(F_1) & \rightarrow 0   \\
    &  \downarrow &  & \downarrow &  &  \downarrow  &  & \downarrow   &     \\
0 \rightarrow &  R_4 & \rightarrow  & J_4(E) & \rightarrow &  J_2(F_0) &  \rightarrow &  F_1 &  \rightarrow 0             \\ 
   &  &  & \downarrow &  &  \downarrow  &  &  \downarrow   & \\
   &  &  &  0  && 0  && 0 &
\end{array}   \]
with dimensions:
\[   \begin{array}{rcccccccl}
    &  0  &  &  0  &  &  0  &  &  0  &   \\
     &  \downarrow &  & \downarrow &  &  \downarrow  &  & \downarrow   &     \\
0 \rightarrow  &  1 & \rightarrow &  6  & \rightarrow &  \fbox { $  8 $} & \rightarrow &  2 & \rightarrow 0  \\
&  \downarrow &  & \downarrow &  &  \downarrow  &  &  \downarrow    &     \\
0 \rightarrow &  4 & \rightarrow  & 21 & \rightarrow & 20 &  \rightarrow &  3 & \rightarrow 0   \\
    &  \downarrow &  & \downarrow &  &  \downarrow  &  & \downarrow   &     \\
0 \rightarrow &  4 & \rightarrow  & 15 & \rightarrow &  12 &  \rightarrow &  1  &  \rightarrow 0             \\ 
   &  &  & \downarrow &  &  \downarrow  &  &  \downarrow   & \\
   &  &  &  0  && 0  && 0 &
\end{array}   \]
The upper symbol sequence is not exact at $S_3 T^* \otimes F_0$ even though the two other sequences are exact on the jet level. As a byproduct we have the exact sequences 
$\forall r\geq 0$:  \\
\[   0 \rightarrow  R_{r + 4} \rightarrow J_{r + 4}(E) \rightarrow  J_{r + 2}(F_0) \rightarrow  J_r(F_1) \rightarrow 0   \] 
Such a result can be checked directly through the identity:   \\
 \[   4 - (r+5)(r+6)/2 + 2(r+3)(r+4)/2 - (r+1)(r+2)/2 = 0   \]
We obtain therefore the formally exact sequence we were looking for, namely:  \\
\[  0 \rightarrow \Theta \rightarrow E \underset 2{\stackrel{{\cal{D}}}{\longrightarrow}}  F_0  \underset 2{\stackrel{{\cal{D}}_1}{\longrightarrow}}  F_1 \rightarrow 0   \]
The {\it surprising fact} is that, in this case $ad({\cal{D}}) $ generates the CC of $ad({\cal{D}}_1) $. Indeed, multiplying by the Lagrange multiplier test function $\lambda$ and integrating by parts, we obtain the second order operator $ \lambda \rightarrow (- d_{22} \lambda = {\mu}^1, d_{12} \lambda - a d_1 \lambda = {\mu}^2 ) $  and thus 
$ - a^2 d_1 \lambda = d_1 {\mu}^1 + d_2 {\mu}^2  + a {\mu}^2 $. Substituting, we finally get the only second order CC  $d_{12} {\mu}^1 + d_{22}{\mu}^2 - a d_1 {\mu}^1=0$.  \\                                                                                                 

In the differential module framework over the commutative ring $ D= K[d_1,d_2]$ of differential operators with coefficients in the trivially differential field $K= \mathbb{Q}(a)$, we have the free resolution:
\[         0 \rightarrow  D  \underset 2{\stackrel{{\cal{D}}_1}{\longrightarrow}} D^2   \underset 2{\stackrel{{\cal{D}}}{\longrightarrow}} D \rightarrow M \rightarrow 0 \]
of the differential module $ M $ with Euler-Poincar\'{e} characteristic $rk_D(M) = 1 - 2 + 1 = 0$. We recall that $R= R_{\infty} = hom_K(M,K)$ is a differential module for the Spencer operator $d: R \rightarrow T^* \otimes R : R_{q+1}  \rightarrow T^* \otimes R_q$ (See [15, 19] for more details).   \\

The two following examples will show how the differential extension modules may depend on the Vessiot structure constants.  \\

\noindent
{\bf EXAMPLE 2.2}: With $m=n=2, q=1, K=\mathbb{Q}<\omega>$ and $\omega=(\alpha,\beta)$ with $\alpha\in T^*,\beta\in {\wedge}^2T^*$, let us consider the Lie operator ${\cal{D}}:T \rightarrow  \Omega: \xi\rightarrow {\cal{L}} (\xi)\omega=(A={\cal{L}}(\xi)\alpha, B={\cal{L}}(\xi)\beta)$. The corresponding first order system:   \\
\[ A_i \equiv {\alpha}_r{\partial}_i{\xi}^r+{\xi}^r{\partial}_r{\alpha}_i=0,  \,\,\, B \equiv \beta {\partial}_r{\xi}^r  +{\xi}^r{\partial}_r \beta =0\]       
is involutive whenever $\beta\neq 0$ and $\fbox{ $d\alpha = c \beta $ } $ where now $d$ is the exterior derivative and $c= cst$. \\
We notice that $\bar{\omega}$ and $\omega$ provide the same system of Lie equations if and only if $\bar{\alpha}= a \alpha, \bar{\beta}=±b \beta$ with $a,b \neq 0$ and thus $\bar{c} = \frac{a}{b} c$, a result showing that the only critical value of $c$ is $c = 0$.  \\
We have the formally exact differential sequence:  \\
\[    0 \rightarrow \Theta \rightarrow T \stackrel{{\cal{D}}}{\longrightarrow}T^*{\times}_X{\wedge}^2T^* \stackrel{{\cal{D}}_1}{\longrightarrow} {\wedge}^2T^* \rightarrow 0  \]
or the resolution:  \\
\[    0 \rightarrow D \stackrel{{\cal{D}}_1}{\longrightarrow} D^3 \stackrel{{\cal{D}}}{\longrightarrow} D^2 \stackrel{p}{\longrightarrow} M  \rightarrow 0  \]
Multiplying $(A_1,A_2,B)$ respectively by $({\mu}^1,{\mu}^2,{\mu}^3)$, we obtain $ad({\cal{D}})$ in the form:  \\
\[ - {\alpha}_1({\partial}_1{\mu}^1+{\partial}_2{\mu}^2) - \beta ({\partial }_1{\mu}^3 - c {\mu}^2)={\nu}^1, \,\, 
   - {\alpha}_2({\partial}_1{\mu}^1+{\partial}_2{\mu}^2) - \beta ({\partial }_2{\mu}^3 + c {\mu}^1)={\nu}^2  \] 
Then, multiplying ${\partial}_1A_2-{\partial}_2A_1- cB$ by $\lambda$, we obtain $ad({\cal{D}}_1)$ as:  \\                       
\[   {\partial}_2\lambda={\mu}^1, \,\, - {\partial}_1 \lambda = {\mu}^2, \,\, - c \lambda={\mu}^3  \]
We have therefore to consider the two cases: \\
\noindent 
$\bullet$ $c=0$: We have the new CC ${\partial}_1{\mu}^1 + {\partial}_2{\mu}^2=0$ and ${\mu}^3=0$. It follows that the torsion module ${ext}^1(M) \neq 0$ is generated by the residue of ${\mu}^3={\nu}' $ because $\alpha \neq 0$ and we may thus suppose that ${\alpha}_1\neq 0$. As for ${ext}^2(M)$, this torsion module is just defined by the system ${\partial}_2\lambda=0, {\partial}_1\lambda=0$ for $\lambda$ and thus ${ext}^2(M)\neq 0$.  \\
\noindent
$\bullet$ $c\neq 0$: We must have the new CC:  \\
\[  {\partial}_1{\mu}^3 -  c{\mu}^2=0, {\partial}_2{\mu}^3 + c{\mu}^1=0\Rightarrow {\partial}_1{\mu}^1+{\partial}_2{\mu}^2=0  \]
It follows that ${ext}^1(M)$ is now generated by the residue of ${\partial}_1{\mu}^1 + {\partial}_2{\mu}^2={\nu}'$. Finally, $ker(ad({\cal{D}}_1))$ is defined by $\lambda=0$ and thus ${ext}^2(M)=0$. \\
Hence, both ${ext}^1(M) $ and ${ext}^2(M)$ highly depend on the Vessiot structure constant $c$.  \\

\noindent
{\bf EXAMPLE 2.3}: ({\it Contact transformations}) \\
 With $m=n=3, q=1, K=\mathbb{Q}(x^1,x^2,x^3)$  or simply $\mathbb{Q}(x)$, we may introduce the $1$-form $\alpha=dx^1-x^3dx^2 \in T^*$ and consider the system of finite Lie equations defined by $j_1(f)^{-1}(\alpha)=\rho(x) \alpha$. Eliminating the factor $\rho$ and linearizing at the $q$-jet of the identity, we obtain a first order system $R_1 \subset J_1(T) $ made by ${\eta}^1$ and ${\eta}^2$ below which is not even formally integrable:   \\
\[  \left\{ \begin{array}{lcr}
 {\partial}_3{\xi}^1 - x^3 {\partial}_3{\xi}^2  & =  & {\eta}^2  \\
 {\partial}_2{\xi}^1 - x^3 {\partial}_2{\xi}^2 +x^3 {\partial}_1{\xi}^1 - (x^3)^2{\partial}_1{\xi}^2 - {\xi}^3 & =  &   {\eta}^1
\end{array}  \right.  \fbox{$ \begin{array}{lll}
1 & 2 & 3  \\
1 & 2 & \bullet
\end{array} $ }  \] 
with one equation of class $3$ and one equation of class $2$. Strikingly, the symbol $g_1$ is involutive with characters ${\alpha}^1_1=3, {\alpha}^2_1=2, {\alpha}^3_1=2$, a result leading to $dim(g_1)= 3 + 2 + 2=7, dim(g_2)= 3 +(2\times 2) + (3 \times 2) = 13$ and more generally $dim(g_{r+1} ) = 3 + 2(r+1) + (r+1)(r+2) = r^2 + 5 r + 7, \forall r\geq 0$. However, the system $R_1$ is not formally integrable and thus not involutive. The PP procedure brings ${\eta}^3 = {\partial}_2{\eta}^2 -  {\partial}_3{\eta}^1+ x^3 {\partial}_1{\eta}^2 $ below and we get the involutive system $R^{(1)}_1 \subset R_1$ of infinitesimal Lie equations having the same solutions, which is already in $\delta$-regular coordinates:  \\
\[  \left\{ \begin{array}{lcr}
{\partial}_3{\xi}^3 + {\partial}_2 {\xi}^2 - {\partial}_1{\xi}^1+ 2 x^3 {\partial}_1{\xi}^2 & = &  {\eta}^3  \\
{\partial}_3{\xi}^1 - x^3 {\partial}_3{\xi}^2  & =   & {\eta}^2 \\
 {\partial}_2{\xi}^1 - x^3 {\partial}_2{\xi}^2 +x^3 {\partial}_1{\xi}^1 - (x^3)^2{\partial}_1{\xi}^2 - {\xi}^3 & =  & {\eta}^1
\end{array}  \right.  \fbox{$ \begin{array}{lll}
1 & 2 & 3  \\
1 & 2 & 3  \\
1 & 2 & \bullet
\end{array} $ }  \]
with two equations of class $3$, one equation of class $2$ and thus one CC of order $1$ described by the trivially involutive first order operator 
${\cal{D}}_1: \eta \rightarrow {\partial}_3{\eta}^1- {\partial}_2{\eta}^2 - x^3 {\partial}_1{\eta}^2 + {\eta}^3 = \zeta$. 

The link existing between these two approaches is not evident as it highly depends on the fact that ${\rho}_r(R^{(1)}_1)  = R^{(1)}_{r+1}$, according to the (quite difficult) theorem 2.4.5 of (See [7], p 70) that must be compared to [5, 26]. Setting $\Phi: J_1(T) \rightarrow J_1(T) / R^{(1)}_1 = F_0 $ with $dim(F_0)=3$ and ${\Phi}': J_1(T) \rightarrow J_1(T) / R_1=F'_0$ with $dim(F'_0)=2$, the inclusion $R^{(1)}_1\subset R_1$ induces an epimorphism $F_0 \rightarrow F'_0 \rightarrow 0$ and we have the commutative diagram when $r\geq 2$ and $F_1=coker({\rho}_1(\Phi)) $:  \\
\[  \begin{array}{rcccccccl}
    &  0  &  &  0  &  &  0  &  &   &   \\
     &  \downarrow &  & \downarrow &  &  \downarrow  &  &   &     \\
0 \rightarrow  &  g_{r+1}  & \rightarrow & S_{r+1}T^*\otimes T & \stackrel{{\sigma}_r({\Phi}')}{\longrightarrow} &  S_rT^*\otimes F'_0 & &   &  \\
&  \downarrow &  & \downarrow &  &  \downarrow  &  &    &     \\
0 \rightarrow &  R_{r+1} & \rightarrow  & J_{r+1}(T) & \stackrel{{\rho}_r({\Phi}')}{\longrightarrow}  &  J_r(F'_0) &  \rightarrow &   0 &   \\
    &  \downarrow &  & \downarrow &  &  \downarrow  &  &  &     \\
0 \rightarrow &  R^{(1)}_r & \rightarrow  & J_r(T) & \stackrel{{\rho}_{r-1} (\Phi)}{\longrightarrow}  &  J_{r-1}(F_0) &  \rightarrow &  J_{r-2}(F_1)  & \rightarrow  0 \\ 
   & \downarrow &  & \downarrow &  &   &  &  & \\
   & 0 &  &  0  &  &  &. &   &
\end{array}  \]
The operator ${\cal{D}}'={\Phi }'\circ j_1$ is formally surjective, a snake chase proves that the columns are exact and the bottom sequence is exact as it is just providing the formally exact Janet sequence:
\[   0 \rightarrow \Theta \rightarrow T \underset 1 {\stackrel{{\cal{D}}}{\longrightarrow}} F_0 \underset 1 {\stackrel{{\cal{D}}_1}{\longrightarrow}} F_1 \rightarrow 0 \] 
Hence, it remains to check that the central row is also exact. For this, we notice that:  \\
\[ dim(R^{(1)}_r) = (r+1)(r+2)(r+3)/2 - r(r+1)(r+2)/2 + r (r-1)(r+1)/6 = (r+1)(r^2 + 8 r + 18)  \]
\[ dim(R_{r+1}) = (r+2)(r+3)(r+4)/2 - (r+1)(r+2)(r+3)/3 = (r+2)(r+3)(r+10)/6 \]
and check that $dim(R_{r+1})  - dim(R^{(1)}_r) = dim(g_{r+1}) $, a result leading to the sequence:  \\
\[  0 \rightarrow \Theta \rightarrow T \underset 1{\stackrel{{\cal{D}}'}{\longrightarrow }} F'_0 \rightarrow 0 \]
\[  \hspace{1cm} ({\xi}^1,{\xi}^2,{\xi}^3) \stackrel{{\cal{D}}' }{\longrightarrow} ({\eta}^1,{\eta}^2)  \rightarrow 0 \]
Now, it is well known that this {\it contact} operator ${\cal{D}}$ allowing to define the contact module $M$ admits an injective parametrization by one arbitrary potential function $\phi$ as follows:  \\
\[  {\cal{D}}_{-1}: \phi \rightarrow  ( - x^3 {\partial}_3\phi + \phi={\xi}^1, \,- {\partial}_3\phi={\xi}^2, \, {\partial}_2\phi+x^3{\partial}_1 \phi={\xi}^3 ) \,\,\, \Rightarrow \,\,\,  {\xi}^1 - x^3 {\xi}^2 = \phi \,\,\,  (lift) \]
and thus $M\simeq D$. We have obtained the following locally exact differential sequence ([8]):  \\
\[  0 \rightarrow \phi \stackrel{{\cal{D}}_{-1}}{\longrightarrow } ({\xi}^1,{\xi}^2,{\xi}^3) \stackrel{{\cal{D}}}{\longrightarrow} ({\eta}^1,{\eta}^2,{\eta}^3)  \stackrel{{\cal{D}}_1}{\longrightarrow} \zeta \rightarrow 0 \]
which is {\it not} a Janet sequence because the first operator is {\it not} involutive. \\
We obtain therefore a resolution of $M$ [13]: \\
 \[    0 \rightarrow D \stackrel{{\cal{D}}_1}{\longrightarrow} D^3 \stackrel{{\cal{D}}}{\longrightarrow } D^3 \rightarrow M \rightarrow 0   \]
but also a resolution which is {\it not} strictly exact [13] because $R_1$ is not even formally integrable:  \\
\[       0 \rightarrow D^2 \stackrel{{\cal{D}}'}{\longrightarrow } D^3 \rightarrow  M \rightarrow 0 \hspace{1cm} \Rightarrow \hspace{1cm} M \oplus D^2 \simeq D^3 \]
As $M$ is free and thus projective, the respective adjoint sequences are also exact. Applying $hom_D(\bullet,D) $, we obtain therefore a split exact sequence of free {\it right} (care) differential modules. Passing from right to left differential modules by ususing the {\it side changing procedure}   $N_D \rightarrow N={ }_D N=hom_K({\wedge}^nT^*, N_D)$, it follows that the adjoint sequence is exact too, though not strictly exact. As such a result does not depend on the differential sequence used, according to a (difficult) general theorem in homological algebra, we shall prove it directly through a (delicate) technical explicit computation for the first of the two previous ones and let the reader prove it similarly for the other.  \\

For this, let us multiply $\zeta$ by a test function $\lambda$ and integrate by parts in order to obtain: 
\[  ad({\cal{D}}_1) : \lambda \rightarrow \left\{ \begin{array}{lcr}
 - {\partial}_3 \lambda & = &  {\mu}^1  \\
{\partial}_2   \lambda + x^3 {\partial}_1 \lambda  & =   & {\mu}^2 \\
   \lambda   & =  & {\mu}^3
\end{array}  \right. \]
This operator is injective but not involutive as the corresponding system $R_1 \subset J_1(\lambda)$ is far from being even formally integrable. In a coherent way with the standard second Spencer sequence [   ], the corresponding involutive operator is $j_1$ as follows, with 6 first order CC for itself but first {\it and} second order for $\mu$ after substitution, that is a second order system which is not FI. \\
\[ j_1: \lambda \rightarrow   \left\{ \begin{array}{lcr}
{\partial}_3    \lambda & = & - {\mu}^1  \\
{\partial}_2 \lambda  & =   & {\mu}^2 - x^3 {\partial}_1 {\mu}^3      \\
{\partial}_1 \lambda  & = & {\partial}_1 {\mu}^3  \\
   \lambda & =  & {\mu}^3
\end{array}  \right.  \fbox{$ \begin{array}{lll}
1 & 2 & 3  \\
1 & 2 & \bullet  \\
1 & \bullet & \bullet. \\
\bullet & \bullet  & \bullet
\end{array} $ }  \]
\noindent
$\bullet$ 1  {\it identity to zero}  ${\partial}_1 {\mu}^3 - {\partial}_1{\mu}^3=0$.  \\
$\bullet$ 2  first order CC, namely $ - {\nu}^3 \equiv {\partial}_3 {\mu}^3 + {\mu}^1=0$, \,  $ - ( {\nu}^2 + x^3 {\nu}^1) \equiv {\partial}_2 {\mu}^3 + x^3 {\partial}_1 {\mu}^3 - {\mu}^2=0$ which are differentially independent because, when ${\mu}^3$ is given arbitrarily, the first is providing ${\mu}^1$ while the second is providing ${\mu}^2$. \\
$\bullet$ 2  second order CC that are just prolongations of the two previous first order ones. \\
$\bullet$ 1 second order CC $ {\partial}_3 {\mu}^2 - x^3 {\partial}_{13} {\mu}^3 - {\partial}_1 {\mu}^3 + {\partial}_2 {\mu}^1=0$. \\
Using the previous second order CC ${\partial}_{13} {\mu}^3 + {\partial}_1 {\mu}^1=0$, we finally get the {\it new} first order CC $ - {\nu}^1 \equiv {\partial}_3 {\mu}^2 + x^3 {\partial}_1 {\mu}^1 + {\partial}_2 {\mu}^1 - {\partial}_1 {\mu}^3=0$ which is a differential consequence of the two previous first order CC because we have ({\it by chance} !) the involutive system:  \\
\[  \left\{ \begin{array}{lcl}
{\partial}_3{\mu}^3 + {\mu}^1 & = & - {\nu}^3  \\
{\partial}_3{\mu}^2 + x^3 {\partial}_1 {\mu}^1 + {\partial}_2 {\mu}^1 - {\partial}_1 {\mu}^3  & =   & - {\nu}^1 \\
 {\partial}_2{\mu}^3 + x^3 {\partial}_1 {\mu}^3 - {\mu}^2 & =  & - ({\nu}^2 + x^3 {\nu}^1)
\end{array}  \right.  \fbox{$ \begin{array}{lll}
1 & 2 & 3  \\
1 & 2 & 3  \\
1 & 2 & \bullet
\end{array} $ }  \]
providing the only first order CC:  \\
\[ ad({\cal{D}}_{-1}): ({\nu}^1, {\nu}^2, {\nu}^3) \rightarrow  x^3 {\partial}_3 {\nu}^1 + 2 {\nu}^1  + {\partial}_3 {\nu}^2 - {\partial}_2 {\nu}^3 - x^3 {\partial}_1 { \nu}^3 = \theta  \]
As a byproduct, we can find the formally exact differential sequence:   \\
\[    0 \leftarrow  \theta \stackrel{ad({\cal{D}}_{-1})}{\longleftarrow} ({\nu}^1,{\nu}^2, {\nu}^3) \stackrel{ad({\cal{D}}) }{\longleftarrow} ({\mu}^1, {\mu}^2, {\mu}^3) \stackrel{ad({\cal{D}}_1) }{\longleftarrow} \lambda \leftarrow 0  \] 
which is {\it exactly} the adjoint of the first differential sequence we have provided.  \\ 

Coming back to the Vessiot structure equations, we notice that $\alpha$ is not invariant by the contact Lie pseudogroup and cannot be considered as an associated geometric object. We have shown in ([8], p 684-691) that the corresponding geometric object is a $1$-form density $\omega$ leading to the system of infinitesimal Lie equations in Medolaghi form:  \\
\[            {\Omega}_i\equiv ( {\cal{L}}(\xi)\omega)_i \equiv {\omega}_r{\partial}_i{\xi}^r - \frac{1}{2}{\omega}_i{\partial}_r{\xi}^r + {\xi}^r{\partial}_r{\omega}_i=0  \]
that may be also written as:  \\
\[    {\partial}_i({\omega}_r{\xi}^r) - \frac{1}{2} {\omega}_i{\partial}_r{\xi}^r + {\xi}^r({\partial}_r{\omega}_i - {\partial}_i{\omega}_r)={\Omega}_i\]
and to the {\it only} Vessiot structure equations, still not known today:  \\
\[   \fbox{  $   {\omega}_1({\partial}_2{\omega}_3- {\partial}_3{\omega}_2) +   {\omega}_2({\partial}_3{\omega}_1- {\partial}_1{\omega}_3) + 
 {\omega}_3({\partial}_1{\omega}_2 - {\partial}_2{\omega}_1) = c  $ }  \]
with the {\it only} structure constant $c$. In the present contact situation, we may choose 
$\omega=(1, -x^3,0)$ and get $c=1$ but we may also choose $\omega=(1,0,0)$ and get $c=0$, these two choices both bringing an involutive system. Let us prove that the situation becomes completely different with the new system:  \\
\[  -2{\Omega}_1\equiv  {\partial}_3{\xi}^3 + {\partial}_2{\xi}^2 - {\partial}_1{\xi}^1=0,\,\, {\Omega}_2\equiv {\partial}_2{\xi}^1=0, \,\, {\Omega}_3\equiv {\partial}_3{\xi}^1=0   \]
having the only CC $ d_2{\Omega}_3 - d_3 {\Omega}_2=0 $. \\
Multilying the three previous equations by the three test functions $\mu$, the only CC by the test function $\lambda$ and integrating by parts, we get the adjoint operators:    \\
\[    0= {\mu}^1, \,\,  {\partial}_3\lambda= {\mu}^2, \,\, - {\partial}_2\lambda={\mu}^3   \]
\[ {\partial}_1{\mu}^1 - {\partial}_2{\mu}^2 - {\partial}_3{\mu}^3= {\nu}^1, \,\, - {\partial}_2{\mu}^1={\nu}^2, \,\, - {\partial}_3{\mu}^1={\nu}^3 \]
It follows that $0\neq D{\xi}^1 =t(M) \subset M$ with a strict inclusion and ${ext}^1(M)\neq 0$. Similarly, $ker(ad({\cal{D}}_1))$ is defined by ${\partial}_2\lambda=0, {\partial}_3\lambda=0$ and thus ${ext}^2(M)\neq 0$.  \\
Our problem will be now to construct and compare the differential sequences:   \\
\[   \phi \stackrel{{\cal{D}}_{-1}}{\longrightarrow} \xi \stackrel{{\cal{D}}}{\longrightarrow} \Omega \stackrel{{\cal{D}}_1}{\longrightarrow}C \]  
\[   \theta \stackrel{ad({\cal{D}}_{-1})}{\longleftarrow} \nu \stackrel{ad({\cal{D}})}{\longleftarrow} \mu \stackrel{ad({\cal{D}}_1)}{\longleftarrow} \lambda  \]                                                                                                                                                                                                                                                      
For this, linearizing the only Vessiot structure equation, we get the CC operator ${\cal{D}}_1$ and the corresponding system ${\cal{D}}_1 \Omega= C$ in the form:  \\
\[   {\omega}_1({\partial}_2{\Omega}_3- {\partial}_3{\Omega}_2) +   {\omega}_2({\partial}_3{\Omega}_1- {\partial}_1{\Omega}_3) +  {\omega}_3({\partial}_1{\Omega}_2  - {\partial}_2{\Omega}_1) \]
\[+ ({\partial}_2{\omega}_3- {\partial}_3{\omega}_2) {\Omega}_1
  + ({\partial}_3{\omega}_1- {\partial}_1{\omega}_3) {\Omega}_2+  ({\partial}_1{\omega}_2- {\partial}_2{\omega}_1) {\Omega}_3= C   \]
Multiplying on the left by the test function $\lambda$ and integrating by parts, we get the operator $ad({\cal{D}}_1)$ in the form:  \\
\[  \left\{   \begin{array}{lccl}
{\Omega}_1 \rightarrow  & {\omega}_3{\partial}_2\lambda - {\omega}_2 {\partial}_3\lambda + 2({\partial}_2{\omega}_3 - {\partial}_3 {\omega}_2)\lambda & = {\mu}^1  \\
{\Omega}_2 \rightarrow  & {\omega}_1{\partial}_3\lambda - {\omega}_3 {\partial}_1\lambda + 2({\partial}_3{\omega}_1 - {\partial}_1 {\omega}_3)\lambda & = {\mu}^2  \\
{\Omega}_3 \rightarrow  & {\omega}_2{\partial}_1\lambda - {\omega}_1 {\partial}_2\lambda + 2({\partial}_1{\omega}_2 - {\partial}_2 {\omega}_1)\lambda & = {\mu}^3  \\
\end{array} \right.  \]
We obtain therefore the crucial formula $2c \,\lambda = {\omega}_i{\mu}^i$ showing how the previous sequences are {\it essentially} depending on the Vessiot structure constant $c$. Indeed, if $c\neq 0$, then $\mu=0 \Rightarrow \lambda =0$ and the operator $ad({\cal{D}}_1)$ is injective. This is the case when $\omega =(1,-x^3,0)\Rightarrow c=1  \Rightarrow \lambda =0$. On the contrary, if $c=0$, then the operator $ad({\cal{D}}_1)$ may not be injective as can be seen by choosing $\omega=(1,0,0)$. Indeed, in this case we get a kernel defined by ${\partial}_3\lambda=0, {\partial}_2\lambda=0$.  \\
Finally, in order to exhibit the generating CC of $ad({\cal{D}}_1)$ when $c\neq 0$, we just need to substitute $\lambda=(1/2c){\omega}_i{\mu}^i$ in the previous equations $ad({\cal{D}}_1)\lambda=\mu$. On the other side, multiplying the equations ${\cal{D}}\xi=\Omega$ by test functions ${\mu}^i$ and integrating by parts, we get $ad({\cal{D}})\mu=\nu$ in the form:  \\
\[ \left\{  \begin{array}{lccl}
{\xi}^1 \rightarrow & \,\,  & - {\partial}_i({\omega}_1{\mu}^i) + \frac{1}{2}{\partial}_1({\omega}_i{\mu}^i)+
({\partial}_1{\omega}_i){\mu}^i&={\nu}^1    \\
{\xi}^1 \rightarrow & \,\,  & - {\partial}_i({\omega}_2{\mu}^i) + \frac{1}{2}{\partial}_2({\omega}_i{\mu}^i)+
({\partial}_2{\omega}_i){\mu}^i & = {\nu}^2   \\
{\xi}^1 \rightarrow & \,\,  & - {\partial}_i({\omega}_3{\mu}^i) + \frac{1}{2}{\partial}_3({\omega}_i{\mu}^i)+
({\partial}_3{\omega}_i){\mu}^i & = {\nu}^3
\end{array}  \right. \]
We let the reader check, as a tricky exercise, that we have indeed $ad({\cal{D}})\circ ad({\cal{D}}_1)\equiv 0$ and it remains to prove that $ad({\cal{D}})$ generates the CC of $ad({\cal{D}}_1)$. It is in such a situation that we can measure the usefulness of homological algebra and we only prove this result directly when $\omega=(1,-x^3,0) \Rightarrow c=1$. In this case, the kernel of $ad({\cal{D}})$ is easily seen to be defined by:  \\
\[ \left\{ \begin{array}{ll}
  \frac{1}{2}{\partial}_1{\mu}^1 +{\partial}_2{\mu}^2 + {\partial}_3{\mu}^3+\frac{1}{2} x^3{\partial}_1{\mu}^2 & =0  \\
  - (x^3)^2{\partial}_1{\mu}^2 + x^3{\partial}_1{\mu}^1+{\partial}_2{\mu}^1-x^3{\partial}_2{\mu}^2+2{\mu}^3 & =0   \\
  {\partial}_3{\mu}^1 - x^3 {\partial}_3{\mu}^2 - 3 {\mu}^2  &  =0
  \end{array}  \right.  \]
while, setting now $2\lambda={\mu}^1-x^3{\mu}^2$ and substituting, the CC of $ad({\cal{D}}_1)$ seem to be only defined by the two PD equations:  \\
\[ \left\{ \begin{array}{ll}
{\partial}_3{\mu}^1 - x^3 {\partial}_3{\mu}^2 - 3 {\mu}^2   &  =0  \\
{\partial}_2{\mu}^1 - x^3 {\partial}_2{\mu}^2 - (x^3)^2 {\partial}_1{\mu}^2 +x^3 {\partial}_1{\mu}^1+ 2{\mu}^3  &  =0
\end{array} \right.   \]
The strange fact is that {\it such a system is not formally integrable} and one has to differentiate the second PD equation with respect to $x^3$ and substract the first PD equation differentiated with respect to $x^2$ in order to get the additional PD equation:   \\
\[       {\partial}_3{\mu}^3 +{\partial}_2{\mu}^2 + \frac{1}{2}x^3 {\partial}_1{\mu}^2+\frac{1}{2}{\partial}_1{\mu}^1=0     \]
and find an isomorphic involutive system, a result showing that the differential sequence and its formal adjoint are both formally exact though {\it not} strictly exact. We conclude this example with the following striking result [arXiv:1803.09610]:  \\

\noindent
{\bf THEOREM 2.3.1}: The {\it contact differential sequence} and its formal adjoint are both split long exact sequences of free and thus projective modules, if and only if $c\neq 0$. \\  

\noindent
{\it Proof}: As we have just proved that $ad({\cal{D}})$ was generating the CC of $ad({\cal{D}}_1)$, we may look for the CC of $ad({\cal{D}})$ in order to recover the parametrization of ${\cal{D}}$ given at the beginning of this example.  \\
First of all, as proved in [GB2], two geometric objects $\omega$ and $\bar{\omega}$ provide the same system $R_1\subset J_1(T) $ if and only if $\bar{\omega}= a \omega$ for a constant parameter $a=cst$, a result providing $\bar{c}=a^2 c$. Accordingly, the only possible critical value of the Vessiot structure constant is $c=0$. \\
Now, as already noticed, we have ${\omega}_i{\mu}^i=2c\lambda$ and $ad({\cal{D}}_1)$ is injective if and only if $c\neq 0$. It follows that the differential module defined by ${\cal{D}}_1$ is projective and the sequences split if we are able to construct ${\cal{D}}_{-1}$ and to prove that it is an injective operator. For this, changing slightly our previous notations, we notice that the symbol map of $ad({\cal{D}})$ is:  \\
\[    - {\omega}_i d_r {\mu}^r+ \frac{1}{2}{\omega}_r d_i {\mu}^r + ... = {\nu}_i    \]
Using standard notations of classical geometry, we may rewrite it as:   \\
\[    - A \vec{\omega} + \vec{\nabla} . B + ... = \vec{\nu}  \Rightarrow \vec{\omega} \wedge ( \vec{\nabla}. A)  + ... = \vec{\nabla} \wedge \vec{\nu} \Rightarrow \vec{\omega} . (\vec{\nabla} \wedge \vec{\nu})  + ... = 0  \]
Hence, after one prolongation on the symbol level, we get the only CC:  \\ 
\[  {\omega}_1 ( d_3 {\nu}_2 - d_2 {\nu}_3) + {\omega}_2 ( d_1 {\nu}_3 - d_3 {\nu}_1) + {\omega}_3 ( d_2 {\nu}_1 - d_1 {\nu}_2) + ... =  0 \]
After substitution and a tedious computation, one finally obtains:  \\
\[   \begin{array}{rcl}
{\omega}_1 ( d_3 {\nu}_2 - d_2 {\nu}_3) + {\omega}_2 ( d_1 {\nu}_3 -  d_3 {\nu}_1) + {\omega}_3 ( d_2 {\nu}_1 - d_1 {\nu}_2)&  & \\
+2({\partial}_2{\omega}_3 - {\partial}_3 {\omega}_2) {\nu}_1 + 2({\partial}_3{\omega}_1 - {\partial}_1 {\omega}_3) {\nu}_2 + 2({\partial}_1{\omega}_2 - {\partial}_2{\omega}_1) {\nu}_3 & = & \theta    
\end{array}    \]
We may thus obtain ${\cal{D}}_{-1}$ in the form:  \\
\[ \left\{    \begin{array}{lccl}
 {\omega}_2{\partial}_3\phi- {\omega}_3 {\partial}_2\phi + ({\partial}_2{\omega}_3 - {\partial}_3 {\omega}_2)\phi & = {\xi}^1  \\
 {\omega}_3{\partial}_1\phi - {\omega}_1 {\partial}_3\phi + ({\partial}_3{\omega}_1 - {\partial}_1 {\omega}_3)\phi & = {\xi}^2  \\
 {\omega}_1{\partial}_2\phi - {\omega}_2 {\partial}_1\phi + ({\partial}_1{\omega}_2 - {\partial}_2 {\omega}_1)\phi & = {\xi}^3  \\
\end{array}  \right. \]
and this operator is injective whenever $c\neq 0$ because we have the lift ${\omega}_r{\xi}^r=c \phi$.  \\
The situation is similar in arbitrary dimension $n=2p+1$ with $1$-form $\alpha= dx^n -{\sum}^p_{\alpha=1}x^{\bar{\alpha}}dx^{\alpha}$ as we 
have again one Vessiot structure constant $c$ and the injective parametrization :  \\
\[  \phi - x^{\bar{\beta}}\frac{\partial \phi}{\partial x^{\beta}}= {\xi}^n, \hspace{6mm}  - \frac{\partial \phi}{\partial x^{\bar{\alpha}}}={\xi}^{\alpha}, \hspace{6mm}  \frac{\partial \phi}{\partial x^{\alpha}}+x^{\bar{\alpha}} \frac{\partial \phi}{\partial x^n}={\xi}^{\bar{\alpha}} \hspace{4mm}\Rightarrow \hspace{4mm}  \phi=\alpha(\xi)  \]
We have the locally exact split differential sequence:  \\
\[  0  \rightarrow {\wedge}^0T^* \stackrel{{\cal{D}}_{-1}}{\longrightarrow} T \stackrel{\cal{D}}{\longrightarrow} F_0 \stackrel{{\cal{D}}_1}{\longrightarrow} F_1 \stackrel{{\cal{D}}_2}{\longrightarrow} ..... \stackrel{{\cal{D}}_{n-2}}{\longrightarrow} F_{n-2} \rightarrow 0  \]
with $dim(F_r)=n!/(r+2)!(n-r-2)!$ and we refer again to [8, 21] for more details because, when $n\geq  5$, we have to use a $2$-contravariant skewsymmetric tensor density. \\
It follows that contact geometry must be entirely revisited in the light of these new results. \\
\hspace*{12cm}   $ \Box $ \\

\noindent
{\bf EXAMPLE 2.4}: ({\it Macaulay example}). \\
With $n=3, m=1, q=2, K=\mathbb{Q}, D=K[d_1, d_2, d_3] $, let us consider the linear homogeneous second order operator ${\cal{D}}: \xi \rightarrow (d_{33} \xi = {\eta}^1, d_{23} \xi - d_{11} \xi = {\eta}^2, d_{22} \xi = {\eta}^3)$ with corresponding system $R_2 \subset J_2(E)$ which is trivially formally integrable (FI). It is easy to check that $dim(g_2)= 3, dim(g_3)=1, g_4=0$. Hence, $R_2$ and thus $R_3$ are FI but not involutive. Then $R_4$ is trivially involutive with $dim(R_{3+r} ) = 2^n = 8, \forall r \geq 0$ and a full basis of parametric jets is 
$(\xi, {\xi}_1, {\xi}2, {\xi}_3, {\xi}_{11}, {\xi}_{12}, {\xi}_{13}, {\xi}_{111})$. We have proved in [14] that there are three second order CC with only one second order CC. The corresponding formally exact differential sequence is:  \\
\[      0 \rightarrow \Theta \rightarrow \xi \underset 2{\stackrel{{\cal{D}}}{\longrightarrow}} \eta \underset 2{\stackrel{{\cal{D}}_1}{\longrightarrow}} \zeta 
\underset 2{\stackrel{{\cal{D}}_2}{\longrightarrow}} \tau \rightarrow 0 \]
and the minimum free differential resolution of the corresponding differential module $M$ is:  \\
\[            0 \rightarrow D \stackrel{{\cal{D}}_2}{\longrightarrow} D^3 \stackrel{{\cal{D}}_1}{\longrightarrow} D^3 \stackrel{{\cal{D}}}{\longrightarrow } D 
\stackrel{p}{\longrightarrow} M \rightarrow 0 \]
where $p$ is the canonical projection. 
The situation is {\it much more tricky} if we consider now the new third order differential operator ${\cal{D}}' = \Phi \circ j_3: E \stackrel{j_3}{\longrightarrow} J_3(E) \stackrel{\Phi}{\longrightarrow} J_3(E)/R_3$ with $R_3 = {\rho}_1(R_2)$. Indeed, it is easy to check that the sequence:  \\
\[ 0 \rightarrow {\wedge}^2 T^* \otimes g_3 \stackrel{\delta}{\longrightarrow} {\wedge}^3 T^* \otimes g_2 \rightarrow 0 \]
is exact because both bundles have the same dimension equal to $3$. It follows that $g_3$ is $3$-acyclic and it follows that the $21$ CC are described by an operator of order $1$. However, the corresponding system has a symbol which is not $2$-acyclic. We have proved in [14] that the next CC are of order $2$, and so on. By chance, it happens that the next operator ${\cal{D}}'_2$ is involutive and we may start a Janet sequence with first order involutive operators ${\cal{D}}'_3, {\cal{D}}'_4, {\cal{D}}'_6$ in the differential sequence:  \\
\[     0 \rightarrow \Theta \rightarrow 1 \underset 3{\stackrel{ {\cal{D}}'} {\longrightarrow}} 12 \underset 1{\stackrel{{\cal{D}}'_1}{\longrightarrow}} 21 \underset 2{\stackrel{{\cal{D}}'_2}{\longrightarrow}} 46 \underset 1{\stackrel{{\cal{D}}'_3}{\longrightarrow}}  72 \underset 1{\stackrel{{\cal{D}}'_4}{\longrightarrow}} 48 \underset 1{\stackrel{{\cal{D}}'_5}{\longrightarrow}} 12   \rightarrow 0         \]
in which we have specified the dimensions of the bundles and the orders of the successive operators. We finally notice again that $rk_D(M) = 1 - 12 + 21 - 46 + 72 - 48 + 12 = 0$.  \\         
We invite the reader to consider the similar example proposed by F. S. Macaulay in $1916$ ([6], p 79) when $n=2$. Indeed, if we consider the second order system $d_{22}\xi=0, d_{12}\xi - d_{11}\xi=0$, we have $dim(g_2)=1, g_3=0$ and $dim(R_2)= dim(R_3)= 2^n=4$ with the basis of parametric jets $(\xi, {\xi}_1, {\xi}_2, {\xi}_{11})$ . In this case, $R_2$ is FI but not involutive while $R_3$ is trivially involutive with Janet sequence $0 \rightarrow \Theta \rightarrow 1 \underset 3{\stackrel{}{\longrightarrow}}  6 \underset 1{\stackrel{}{\longrightarrow }} 7 \underset 1{\stackrel{}{\longrightarrow}} 2 \rightarrow 0$ and $1 - 6 + 7 - 2 = 0$.  \\ \\

\noindent
{\bf 3) KERR METRIC} \\

Let us recall a few facts from Riemannian geometry. A metric $\omega = ({\omega}_{ij}) \in S_2T^*$ with $det(\omega) \neq 0$ is providing the Christoffel symbols 
$\gamma = ({\gamma}^k_{ij}) $ as geometric objects according to the forgotten work of E. Vessiot in 1903 [7, 24, 27]. The Riemann tensor ${\rho}^k_{l,ij} \in {\wedge}^2 T^* \otimes g_1 \subset {\wedge}^2 T^* \otimes T^* \otimes T $ is a section of the Spencer $\delta$-chomology vector bundle $H^2(g_1)$ because it is also killed by the Spencer surjective map $\delta: {\wedge}^2 T^* \otimes g_1 \rightarrow {\wedge}^3 T^* \otimes T^* \otimes T$ and has thus $n^2 (n-1)^2/4 - n^2 (n-1)(n-2)/6=n^2(n^2-1)/12$ components because $dim(g_1)=n(n-1)/2$ and $g_2=0$.  Needless to say that these definitions are far from the ones that can be found in any place (books or papers) dealing with GR ([12, 15]).  \\
Now, the linearization $\Omega =({\Omega}_{ij})  \in S_2 T^*$ of $\omega$ induces a linearization $\Gamma = {\Gamma}^k_{ij} \in S_2T^* \otimes T$ and a linearization $R=(R^k_{l,ij}) \in {\wedge}^2 T^* \otimes T$.
With more details, we have:  \\
\[  {\rho}^k_{l,ij}= {\partial}_i{\gamma}^k_{lj} - {\partial}_j{\gamma}^k_{li} + 
{\gamma}^r_{lj} {\gamma}^k_{ri} - {\gamma}^r_{li} {\gamma}^k_{rj}   \]
and thus, because $\Gamma \in S_2T^*\otimes T$ is a tensor::   \\
\[ \begin{array}{rcl}
R^k_{l,ij}& = & d_i{\Gamma}^k_{lj} - d_j{\Gamma}^k_{li} + 
{\gamma}^r_{lj} {\Gamma}^k_{ri} - {\gamma}^r_{li} {\Gamma}^k_{rj}   +
{\gamma}^k_{ri} {\Gamma}^r_{lj} - {\gamma}^k_{rj} {\Gamma}^r_{li}   \\
& = &  (d_i{\Gamma}^k_{lj} - {\gamma}^r_{li} {\Gamma}^k_{rj}   + {\gamma}^k_{ri} {\Gamma}^r_{lj} ) - ( d_j{\Gamma}^k_{li} - {\gamma}^r_{lj} {\Gamma}^k_{ri}   + 
{\gamma}^k_{rj} {\Gamma}^r_{li})  \\
 & = &  {\nabla}_i {\Gamma}^k_{lj} - {\nabla}_j {\Gamma}^k_{li} 
\end{array}   \]
by introducing the covariant derivative $\nabla$. We recall that ${\nabla}_r{\omega}_{ij}=0,\forall r,i,j$ or, equivalently, that $(id, -\gamma): \xi \in T \rightarrow ({\xi}^k, {\xi}^k_i= - {\gamma}^k_{ir}{\xi}^r) \in R_1$ is a $R_1$-connection with ${\omega}_{sj}{\gamma}^s_{ir} + {\omega}_{is} {\gamma}^s_{jr}={\partial}_r{\omega}_{ij}$, a result allowing to move down the index $k$ in the previous formulas (See [17, 18, 22] for more details). \\
We may thus take into account the Bianchi identities implied by the cyclic sums on $(ijr)$   \\
\[  {\beta}_{kl,ijr}\equiv {\nabla}_r {\rho}_{kl,ij} + {\nabla}_i{\rho}_{kl,jr} + {\nabla}_j{\rho}_{kl,ri}=0 \hspace{1cm} \Leftrightarrow \hspace{1cm} 
\beta \equiv \underset {cycl}{\Sigma} ( \partial \rho - \gamma \rho)=0\]
and their respective linearizations $B_{kl,ijr}=0$ as described below. Both $\beta$ and $B$ are sections of the Spencer cohomogy vector bundle $H^3(g_1)$ defined by the short exact sequence:  \\
\[   0 \rightarrow H^3(g_1)  \rightarrow {\wedge}^3T^* \otimes g_1 \stackrel{\delta}{\longrightarrow} {\wedge}^4T^* \otimes T \rightarrow 0 \] 
\[   \begin{array}{lcl}
dim(H^3(g_1)) & =  &   (n(n-1)(n-2)/6)(n(n-1)/2) - (n(n-1)(n-2)(n-3)/24)n      \\
                        & =    &    n^2(n^2-1)(n-2)/24
 \end{array}   \]                                       
that is $20$ when $n=4$ because $dim(g_1)=n(n-1)/2 $ and the coboundary bundle is $B^3(g_1)=0$. \\
{\it Such results cannot be even imagined by somebody not aware of the $\delta$-acyclicity} ([7, 9, 11, 12]). \\
We have the linearized cyclic sums of covariant derivatives both with their respective symbolic descriptions, not to be confused with the non-linear corresponding ones:  \\
\[  \begin{array}{rcl}
B_{kl,rij}\equiv  {\nabla}_rR_{kl,ij} + {\nabla}_iR_{kl,jr} + {\nabla}_jR_{kl,ri}= 0 \hspace{1mm}  mod(\Gamma)  & \Leftrightarrow &
\underset{cycl}{\Sigma} ( dR - \gamma R  - \rho \Gamma) =0 \\
    & \Leftrightarrow & B \equiv \underset{cycl}{\Sigma}(\nabla R) =\underset{cycl}{\Sigma}(\rho \Gamma)
    \end{array}  \]
In order to recapitulate these new concepts obtained after one, two or three prolongations, we have successively $ \omega \longrightarrow \gamma \longrightarrow \rho \longrightarrow \beta $  and the respective linearizations $\Omega \longrightarrow \Gamma \longrightarrow R \longrightarrow B $.  \\

The purpose of this section will be to consider the Killing operator ${\cal{D}}: T \rightarrow S_2 T^* : \xi  \rightarrow  {\cal{L}} (\xi)\omega$ and the corresponding Killing system 
$ R_1 \subset J_1(T)$ when $\omega$ is the Kerr metric. In particular, we shall exhibit the successive inclusions $R^{(3)}_1 \subset  R^{(2)}_1 \subset  R^{(1)}_1= R_1 \subset J_1(T)$ with dimensions $ 2 < 4 < 10 = 10 < 20$ by means of elementary combinatorics and diagram chasing exactly like in the previous motivating examples.   \\

We now write the Kerr metric in Boyer-Lindquist coordinates:  \\
\[   ds^2    =   \frac{{\rho}^2 - mr}{{\rho}^2}dt^2 - \frac{{\rho}^2}{\Delta} dr^2  -  {\rho}^2 d{\theta}^2  \\
        - \frac{2a m r sin^2(\theta)}{{\rho}^2} dtd\phi - (r^2+a^2 + \frac{m a^2 r sin^2(\theta)}{{\rho}^2})sin^2(\theta) d{\phi}^2      \]
where $  \Delta= r^2  -mr +a^2 , \,\,  {\rho}^2=r^2 + a^2 cos^2(\theta) $ as usual and we recover the Schwarschild metric when $a=0$. We notice that $t$ or $\phi$ do not appear in the coefficients of the metric. We shall simplify the coordinate system by using the so-called " {\it rational polynomial} " coefficients as follows: \\
 \[ (x^0=t, \, x^1=r, \, x^2=c=cos(\theta), \, x^3=\phi)  \Rightarrow dx^2= - sin(\theta)d\theta \Rightarrow (dx^2)^2=(1-c^2)d\theta^2 \]
We obtain over the differential field $K= \mathbb{Q}(a,m)(t,r,c,\phi)=\mathbb{Q}(a,m)(x)$: \\
\[  ds^2 = \frac{{\rho}^2 - mr}{{\rho}^2} dt^2 - \frac{{\rho}^2}{\Delta} dr^2  -  \frac{{\rho}^2}{1- c^2} dc^2  
 - \frac{2a m r(1- c^2) }{{\rho}^2} dt d\phi -  (1- c^2)( r^2+a^2 + \frac{m a^2 r (1- c^2)}{{\rho}^2}) {d\phi}^2          \]
with now $\Delta = r^2 - mr + a^2$ and $  {\rho}^2 = r^2 + a^2c^2$. For a later use, it is also possible to set ${\omega}_{33}= - (1-c^2)((r^2 + a^2)^2 - a^2 ((1-c^2)(a^2 - mr + r^2))/ (r^2 + a^2c^2)$ and $det(\omega)= - (r^2 + a^2c^2)^2$. \\

Framing the leading derivatives and setting $\xi \partial = {\xi}^i {\partial}_i$, we obtain:   \\
\[ R_1\subset J_1(T) \,\,\,  \left\{  \begin{array}{lcl}
{\Omega}_{33} & \equiv & 2( {\omega}_{33}\fbox{${\xi}^3_3$} + {\omega}_{03} {\xi}^0_3 ) + \xi \partial {\omega}_{33}=0 \\
{\Omega}_{23} & \equiv & {\omega}_{33}\fbox{${\xi}^3_2$} + {\omega}_{03}{\xi}^0_2 + {\omega}_{22}{\xi}^2_3 = 0 \\
{\Omega}_{22} & \equiv & 2 {\omega}_{22}\fbox{${\xi}^2_2$} + \xi \partial {\omega}_{22} =0  \\
{\Omega}_{13} & \equiv  & {\omega}_{33}\fbox{${\xi}^3_1$} + {\omega}_{03}{\xi}^0_1 +{\omega}_{11}{\xi}^1_3 =0 \\
{\Omega}_{12} & \equiv  & {\omega}_{22}\fbox{${\xi}^2_1$} +{ \omega}_{11}{\xi}^1_2 =0  \\
{\Omega}_{11} & \equiv  & 2 {\omega}_{11} \fbox{${\xi}^1_1$} + \xi \partial {\omega}_{11} = 0  \\
{\Omega}_{03} & \equiv  & {\omega}_{33}\fbox{${\xi}^3_0$} + {\omega}_{03}({\xi}^0_0 +{\xi}^3_3)  + {\omega}_{00}{\xi}^0_3 + \xi \partial {\omega}_{03} = 0 \\
{\Omega}_{02} & \equiv  & {\omega}_{22}\fbox{${\xi}^2_0$} + {\omega}_{00}{\xi}^0_2  + {\omega}_{03}{\xi}^3_2 = 0 \\
{\Omega}_{01} & \equiv  & {\omega}_{11}\fbox{${\xi}^1_0$} + {\omega}_{00}{\xi}^0_1 + {\omega}_{03}{\xi}^3_1 =0  \\
{\Omega}_{00} & \equiv  & 2 ({\omega}_{00}\fbox{${\xi}^0_0$} + {\omega}_{03} {\xi}^3_0 ) +\xi \partial {\omega}_{00}=0 
\end{array}  \right.  \]

Looking at the symbol $g_1 \subset T^* \otimes T$, elementary linear combinatorics allow to prove [18, 22]:  \\
\[ {\omega}_{33}\fbox{${\xi}^3_3$} + {\omega}_{03}{\xi}^0_3=0 \,\, mod(\xi), \,\,\, {\omega}_{33} \fbox{${\xi}^3_0$} + 
{\omega}_{00}{\xi}^0_3 = 0 \,\, mod(\xi), \,\,\, 
  {\omega}_{33}\fbox{${\xi}^0_0$} - {\omega}_{03}{\xi}^0_3=0 \,\, mod(\xi)   \] 
Then, multiplying ${\Omega}_{22}$ by ${\omega}_{11}$, ${\Omega}_{11}$ by ${\omega}_{22}$ and adding, we finally obtain:  \\
 \[   2 ({\omega}_{11}{\omega}_{22})({\xi}^1_1 + {\xi}^2_2) +\xi \partial ({\omega}_{11}{\omega}_{22})=0   \]
Using the rational coefficients in the differential field $K=\mathbb{Q}(m,a)(r, c)$, the nonzero components of the corresponding Riemann tensor can be found in textbooks. Among them, we have: \\
\[  \left\{ \begin{array}{rcl} 
  {\rho}_{03,03} & =  & \frac{ mr (1-c^2)(r^2 - mr +a^2)(r^2 - 3a^2c^2)}{2 (r^2 + a^2c^2)^3}  \\  \\  
 {\rho}_{12,12}  & = &  - \frac{mr(r^2 - 3a^2c^2)}{ 2 (1-c^2)(r^2 + a^2c^2)(r^2- mr + a^2)}   \\ \\
 {\rho}_{13,13}  & = &   \frac{ - (1-c^2)mr(r^4 - 2a^2c^2 r^2 + 4a^2r^2 - 2 a^4c^2+ 3a^4- 2a^2mr(1-c^2))(r^2 - 3 a^2c^2)}{2(r^2 + a^2c^2)^3(r^2 - mr +a^2)}  \\ \\
 {\rho}_{23,23}  & = & \frac{mr(2r^4 - a^2c^2 r^2 +5a^2r^2 - a^4c^2 + 3 a^4 - a^2mr(1-c^2))(r^2 - 3 a^2c^2)}{2(r^2 + a^2c^2)^3}   \\ \\
 {\rho}_{01,23}  & =  &  \frac{amc(2r^2 - a^2c^2 + 3 a^2)(3r^2 - a^2c^2)}{ 2 (r^2 + a^2c^2)^3}   \\ \\
 {\rho}_{02,31}  & = & -  \frac{ amc(r^2 - 2 a^2c^2 + 3a^2)(3 r^2 - a^2c^2)}{ 2 (r^2 + a^2c^2)^3}  \\ \\
 {\rho}_{03,12}  & = & - \frac{amc(3r^2 - a^2c^2)}{2(r^2 + a^2c^2)^2} 
\end{array}  \right. \]
As they will be used in the sequel, we also provide a few components vanishing for the S-metric:  \\  
\[   \left \{  \begin{array}{rcl}
 {\rho}_{02,10}  & =  & \frac{3a^2mc(3r^2 - a^2c^2)}{2(r^2 + a^2c^2)^3}  \\  \\
 {\rho}_{02,32}  & =  & \frac{ a m r (3 r^2 - m r + 3 a^2)(r^2 - 3 a^2c^2) }{ 2 (r^2 + a^2c^2)^3}  \\   \\
 {\rho}_{13,23} & =  & - \frac{ 3a^2 m c(1-c^2)(r^2 + a^2)(3 r^2 - a^2c^2)}{2 (r^2 + a^2c^2)^3}  \\   \\  
 {\rho}_{01,13} &  =  &  \frac{amr(1-c^2)(3r^2 + 3a^2 - 2mr)(r^2 - 3a^2c^2)}{ 2 (r^2 + a^2c^2)^3(r^2 -mr+a^2)} 
 \end{array} \right.  \]
We finally exhibit the $6$ vanishing components of the Riemann tensor:  \\
\[      {\rho}_{01,03}=0, {\rho}_{01,12}=0,{\rho}_{02,03}=0, {\rho}_{02,12}=0, {\rho}_{03,13}=0, {\rho}_{03,23}=0, {\rho}_{12,13}=0, {\rho}_{12,23}=0  \] 

Introducing the {\it formal Lie derivative} $R=L({\xi}_1)\rho$ and using the fact that $\rho \in {\wedge}^2T^*\otimes T^* \otimes T^*$ is a tensor, 
the system $R^{(2)}_1\subset R^{(1)}_1 =R_1 \subset J_1(T)$ contains the new equations:  \\
\[  R_{kl,ij}\equiv{\rho}_{rl,ij}{\xi}^r_k + {\rho}_{kr,ij}{\xi}^r_l + {\rho}_{kl,rj}{\xi}^r_i + {\rho}_{kl,ir}{\xi}^r_j + {\xi}^r{\partial}_r{\rho}_{kl,ij}=0 \]
but we have to take into account the ten identities to zero obtained because ${\rho}_{ij}=0$, namely:  \\
\[  R_{ij} \equiv {\rho}_{rj}{\xi}^r_i + {\rho}_{ir}{\xi}^r_J + {\xi}^r{\partial}_r {\rho}_{ij}=0  \]
We check the identity ${\rho}_{01,23} + {\rho}_{02,31} + {\rho}_{03,12} =0 \Rightarrow  R_{01,23} + R_{02,31} + R_{03,12}=0$ and we have:  \\
\[  \left\{  \begin{array}{lcl}
  R_{03,03}  & \equiv  &  2 {\rho}_{03,03} ({\xi}^0_0 + {\xi}^3_3) + \xi \partial {\rho}_{03,03} = 0  \\
  R_{12,12}  & \equiv  & 2 {\rho}_{12,12} ({\xi}^1_1 + {\xi}^2_2) + \xi \partial {\rho}_{12,12} = 0 \\
  R_{01,23}  & \equiv  &  {\rho}_{01,23} ({\xi}^0_0 +{\xi}^1_1 + {\xi}^2_2 + {\xi}^3_3 ) +\xi \partial {\rho}_{01,23}=0  
\end{array} \right. \]
We obtain therefore $ \xi \partial ({\rho}_{12,12}/ ({\omega}_{11}{\omega}_{22})) = 0 $ but we have also $ \xi \partial ( {\rho}_{03,03}{\rho}_{12,12}/det(\omega)) = 0 $.  \\
The following invariants are obtained successively in a coherent way:   \\
\[  {\omega}_{11}{\omega}_{22}= \frac{(r^2 + a^2c^2)^2}{(1-c^2)(r^2 - mr + a^2)} \,\,\, \Rightarrow \,\,\,
\mid {\rho}_{12,12}\mid/({\omega}_{11}{\omega}_{22})= \frac{mr (r^2 - 3 a^2c^2)}{2(r^2+a^2c^2)^3}  \]
Also, as $a\in K$, then ${\rho}_{01,23}$ and ${\rho}_{02,13}$ can be both divided by $a$ and we get the new invariant:  \\
\[  {\rho}_{01,23}/{\rho}_{03,12}= \frac{2r^2-a^2c^2+3a^2}{r^2 + a^2c^2} \]
These results are leading to  \fbox{${\xi}^1=0$},  \fbox{${\xi}^2=0$}, thus to \fbox{${\xi}^1_1=0$}, \fbox{${\xi}^2_2=0$} and ${\xi}^0_0 + {\xi}^3_3=0$ after substitution. In the case of 
the S-metric, we have only $ {\xi}^1=0$ as an equation of zero order. \\
The following relations allow to keep only the four parametric jets $( {\xi}^1_0, {\xi}^2_0, {\xi}^1_3, {\xi}^2_3)$ {\it on the right side}:  \\
\[  \fbox{ $ \left\{  \begin{array}{lcl} 
{\omega}^{11}{\xi}^0_1 + {\omega}^{00} {\xi}^1_0 + {\omega}^{03}{\xi}^1_3=0 ,& \,\,\, &
{\omega}^{11}{\xi}^3_1 + {\omega}^{03} {\xi}^1_0 + {\omega}^{33}{\xi}^1_3 =0  \,\,\, \\
{\omega}^{22}{\xi}^0_2 + {\omega}^{00} {\xi}^2_0 + {\omega}^{03}{\xi}^2_3 =0 , &  \,\,\,  &
{\omega}^{22}{\xi}^3_2 + {\omega}^{03} {\xi}^2_0 + {\omega}^{33}{\xi}^2_3 =0
\end{array}  \right. $ } \]
if we use the fact that $ {\omega}^{03} = - {\omega}_{03}/ ({\omega}_{00}{\omega}_{33} - ({\omega}_{03})^2) $ in the inverse metric.  \\

Taking now into account that $R_{ij} = {\omega}^{rs} R_{ri,sj} \,\,\, mod(\Omega)$, we have proved in [18] that the $6$ second order equations $R_{ij,ij}=0$ are only depending on the $3$ equations $R_{01,01}=0, R_{02,02}=0, R_{01,13}=0$ by using $R_{ii}=0$ and $R_{03}=0$. We obtain in particular the two equations:  \\
\[ \left\{  \begin{array}{lcl}
 \frac{1}{2} R_{01,01} &  \equiv   &    {\rho}_{01,01} ({\xi}^0_0 + {\xi}^1_1) + {\rho}_{01,31}{\xi}^3_0 + {\rho}_{01,02} {\xi}^2_1  = 0  \\ \\
 \frac{1}{2} R_{02,02}  &  \equiv   &  {\rho}_{02,02} ({\xi}^0_0 + {\xi}^2_2) + {\rho}_{02,32}{\xi}^3_0 + {\rho}_{01,02} {\xi}^1_2   = 0   
\end{array}  \right. \]
Using the fact that we have now:  \\
\[ {\omega}_{22}{\xi}^2_1 + {\omega}_{11}{\xi}^1_2=0 \,\,\, \Leftrightarrow \,\,\, {\omega}^{11}{\xi}^2_1 + {\omega}^{22}{\xi}^1_2=0   \]
we may multiply the first equation by ${\omega}^{11}$, the second by ${\omega}^{22}$ and sum in order to obtain:  \\
\[   ({\omega}^{11}{\rho}_{01,01} + {\omega}^{22}{\rho}_{02,02}){\xi}^0_0 + ({\omega}^{11}{\rho}_{01,31} + 
{\omega}^{22}{\rho}_{02,32}){\xi}^3_0 = 0  \]
Using the previous identities for $R_{00}=0$ and $R_{03}=0$, we obtain therefore:  \\
\[    {\omega}^{33} {\rho}_{03,03}{\xi}^0_0 - {\omega}^{03}{\rho}_{03,03}{\xi}^3_0=0 \,\, \Rightarrow \,\, 
{\omega}^{33}{\xi}^0_0 - {\omega}^{03}{\xi}^3_0=0 \,\, \Leftrightarrow \,\, {\omega}_{00}{\xi}^0_0  + {\omega}_{03}{\xi}^3_0=0  \]
As this relation was satisfied $mod(\xi)$, we must introduce the new equation:  \\
\[   R_{01,13} \equiv {\rho}_{01,13} ({\xi}^0_0 + 2 {\xi}^1_1 + {\xi}^3_3) - {\rho}_{13,13} {\xi}^3_0 - {\rho}_{01,01} {\xi}^0_3 + ({\rho}_{01,23} + {\rho}_{02,13} ) {\xi}^2_1 = 0   \]
As we have ${\xi}^0_0 + {\xi}^3_3=0$ and ${\xi}^1_1=0$, we obtain therefore a linear equation of the form:  \\
\[  \fbox { $  {\rho}_{13,13} {\xi}^3_0 + {\rho}_{01,01} {\xi}^0_3 - ({\rho}_{01,23} + {\rho}_{02,13} ) {\xi}^2_1 = 0 $ }   \]
Similarly, we have also:  \\
\[  R_{01,02} \equiv {\rho}_{01,02} ( 2 {\xi}^0_0 + {\xi}^1_1 + {\xi}^2_2 ) - ({\rho}_{01,23} + {\rho}_{02,13} ) {\xi}^3_0 + {\rho}_{01,02} {\xi}^1_2 + {\rho}_{01,02} {\xi}^2_1 = 0   \]
and we obtain therefore a linear equation of the form:  \\
\[  \fbox { $  2 {\rho}_{01,02} {\xi}^0_0 - ({\rho}_{01,23} + {\rho}_{02,13}) {\xi}^3_0 + {\rho}_{01,01} {\xi}^1_2 + {\rho}_{01,02} {\xi}^2_1 = 0 $ }  \]    
 In the case of the S-metric, that is when $a=0$, we obtain respectively ${\xi}^0_3= 0$ and ${\xi}^1_2=0$ as in [18] because ${\xi}^0_0 \simeq {\xi}^0_3$. The previous linear system has thus a rank equal to $2$ and we obtain therefore because ${\xi}^3_0 \simeq {\xi}^0_3, {\xi}^2_1 \simeq {\xi}^1_2$:  \\
 \[   \fbox{${\xi}^0_3=0$},  \fbox{${\xi}^1_2=0$} \,\, \Leftrightarrow \,\,  \fbox{${\xi}^3_0=0$}, \fbox{${\xi}^2_1=0$} , \fbox{${\xi}^0_0=0$}, \fbox{ $ {\xi}^3_3=0 $}  \]

It remains to study the following $4$ linear equations, namely [18]:  \\
\[    R_{01,03} = 0, \,\, R_{03,23} = 0, \,\, R_{03,13} = 0, \,\, R_{02,03} = 0   \] 

\noindent
{\bf THEOREM  3.1}: The rank of the previous system with respect to the $4$ jet coodinates $({\xi}^1_0, {\xi}^2_0, {\xi}^1_3,{\xi}^2_3)$ is equal to $2$, for both the S and K-metrics. We have the two striking identities:  \\
\[  \fbox{ $  R_{03,13} + a(1-c^2) R_{01,03} = 0, \,\,\,\,\,\,   R_{02,03} + \frac{a}{(r^2+a^2)} R_{03,23} = 0 $ } \]  \\

We may define the so-called {\it algebroid brackety}  for sections ${\xi}_q, {\eta}_q \in J_q(T) $ by setting ([7, 8, 9]:  \\
\[    [ {\xi}_q, {\eta}_q ] = \{ {\xi}_{q+1}, {\eta}_{q+1} \} + i(\xi) d {\eta}_{q+1} - i(\eta) d {\xi}_{q+1} \ J_q(T)  \]
In this formula, " $i ( ) $" is the interior multiplication of a $1$-form by a vector and we let the reader check that the right member does not depend on the respective lifts when the 
{\it algebraic bracket} is defined by linearity from the standard bracket of vector fields by the formula:  \\
\[       \{ j_{q+1}(\xi) , j_{q+1} (\eta) \} = j_q ([ \xi, \eta ]) \in J_q(T)  \]

Now, we know that if $R_q\subset J_q(T)$ is a system of infinitesimal Lie equations, then we have the algebroid bracket and its link with the {\it prolongation/projection} (PP) procedure ([7, 8, 14, 15]):  \\
\[  [R_q,R_q]\subset R_q \Rightarrow [R^{(s)}_{q+r}, R^{(s)}_{q+r}] \subset R^{(s)}_{q+r}, \forall q,r,s \geq 0  \]
As $R^{(1)}_1={\pi}^2_1(R_2)=R_1$, it follows that $R^{(2)}_1={\pi}^3_1(R_3)$ is such that $[R^{(2)}_1,R^{(2)}_1]\subset R^{(2)}_1$ with $dim(R^{(2)}_1)= 20-16=4$ because we have obtained a total of $6$ {\it new different} first order equations. Using the first general diagram of the Introduction, we discover that the operator defining $R_1$ has $10+4=14$ CC of order $2$, a result obtained totally independently of the specific GR technical objects (Teukolski scalars, Killing-Yano tensors) introduced in ([1-4]).  \\

We have {\it on sections} ({\it care}) the $16$ (linear) equations $mod(j_2(\Omega))$ of $R^{(2)}_1$ as follows ([22]):\\
\[ R^{(2)}_1 \subset R_1 \subset J_1(T) \, \left\{   \begin{array}{lcl} 
{\xi}^1=0,{\xi}^2=0  &  \Rightarrow & \fbox{ $ {\omega}_{00}{\xi}^0_1 + {\omega}_{03}{\xi}^3_1$} + {\omega}_{11}{\xi}^1_0 =0 ,  \,\, {\xi}^1_1=0, \,\,{\xi}^2_2=0  \\ 
{\xi}^1_2=0  &  \Rightarrow & {\xi}^2_1=0  \\
{\xi}^1_3 + lin({\xi}^1_0,{\xi}^2_0)=0  & \Rightarrow & \fbox{ $ {\omega}_{03}{\xi}^0_1 + {\omega}_{33}{\xi}^3_1$} + {\omega}_{11}{\xi}^1_3 =0   \\
{\xi}^2_3+lin({\xi}^1_0,{\xi}^2_0) =0  & \Rightarrow & \fbox{ $ {\omega}_{00}{\xi}^0_2 +  {\omega}_{03}{\xi}^3_2 $}+ 
{\omega}_{22}{\xi}^2_0 =0 ,  \\
    &  &  \fbox{ $ {\omega}_{03}{\xi}^0_2 +{\omega}_{33}{\xi}^3_2 $}+ {\omega}_{22}{\xi}^2_3=0 \\
{\xi}^0_3=0  & \Rightarrow & {\xi}^3_0=0, \,\, {\xi}^0_0=0, \,\,{\xi}^3_3=0  
\end{array} \right.   \]
The coefficients of the linear equations $lin$ involved depend on the Riemann tensor as in ([22]). Accordingly, we may choose only the $2$ parametric jets $({\xi}^1_0,{\xi}^2_0)$ among $ ({\xi}^1_0, {\xi}^1_3,{\xi}^2_0, {\xi}^2_3)$ to which we must add $({\xi}^0,{\xi}^3)$ {\it in any case} as they are not appearing in the Killing equations. \\
The system is {\it not} involutive because its symbol is finite type but non-zero.   \\ 

Using {\it one more prolongation}, all the {\it sections} ({\it care again}) vanish but ${\xi}^0$ and ${\xi}^3$, a result leading to $dim(R^{(3)}_1)=2$ in a coherent way with the only nonzero Killing vectors $\{ {\partial}_t, {\partial}_{\phi} \}$. We have indeed:  \\
\[   \fbox{$  {\xi}^1_0 =0$} \, ,\,\,\,  \fbox{ $  {\xi}^2_0=0$} \,\,  \Rightarrow  \,\,  {\xi}^1_3=0 , \,\, {\xi}^2_3=0 \,\, 
  \Rightarrow  \,\,  {\xi}^0_1=0 , \,\, {\xi}^3_1=0, \,\,  {\xi}^0_2=0 , \,\, {\xi}^3_2=0 \]

Taking therefore into account that the metric only depends on $(x^1=r,x^2=c=cos(\theta))$ we obtain {\it after three prolongations} the first order system 
$R^{(3)}_1 \subset R^{(2)}_1\subset  R^{(1)}_1 = R_1 \subset J_1(T) $ which is trivially involutive because all the first order jets vanish but $({\xi}^0, {\xi}^3)$.  \\
 
{\it Surprisingly and contrary to the situation found for the S metric}, we have now an involutive first order system with only solutions $({\xi}^0=cst, {\xi}^1=0, {\xi}^2=0, {\xi}^3=cst)$ and notice that $R^{(3)}_1$ does not depend any longer on the parameters $(m,a)\in K$. The difficulty is to know what second members must be used along the procedure met for all the motivating examples. In particular, we have again identities to zero like $d_0{\xi}^1 - {\xi}^1_0=0, d_0{\xi}^2 - {\xi}^2_0=0$ and thus {\it at least} $6$ third order CC coming from the $6$ following components of the Spencer operator, 
namely:  \\
\[ \fbox {$ d_1{\xi}^1 - {\xi}^1_1=0,\,\, d_2{\xi}^1 - {\xi}^1_2=0, \,\,  d_3 {\xi}^1 - {\xi}^1_3 =0, \,\, d_1{\xi}^2 - {\xi}^2_1=0, \,\, d_2{\xi}^2 - {\xi}^2_2=0, \,\, 
d_3 {\xi}^2 - {\xi}^2_3 = 0 $ } \] 
a result that cannot be even imagined from ([1-4]). Of course, proceeding like in the motivating examples, we must substitute in the right members the values obtained from $j_2(\Omega)$ and set for example ${\xi}^1_1= - \frac{1}{2{\omega}_{11}}\xi \partial {\omega}_{11}$ while replacing ${\xi}^1$ and ${\xi}^2$ by the corresponding linear combinations of the Riemann tensor already obtained for the right members of the two zero order equations. \\

Along with the examples, we recall below {\it all} the diagrams that {\it must} be introduced successively:  \\

\[   \begin{array}{rccccccl}
    &    &  &  0  &  &  0  &  &      \\
     &   &  & \downarrow &  &  \downarrow  &  &        \\
     &  0  & \rightarrow & S_2 T^*\otimes T & \stackrel{{\sigma}_1(\Phi)}{\longrightarrow} &  T^*\otimes S_2T^* & \rightarrow  0   \\
&  \downarrow &  & \downarrow &  &  \downarrow  &  &       \\
0 \rightarrow &  R_2& \rightarrow  & J_2(T) & \stackrel{{\rho}_1(\Phi)}{\longrightarrow}  &  J_1(S_2T^*) &  \rightarrow  0  \\
    &  \downarrow &  & \downarrow &  &  \downarrow  &  &     \\
0 \rightarrow &  R_1 & \rightarrow  & J_1(T) & \stackrel{\Phi}{\longrightarrow}  &  S_2T^* &  \rightarrow   0  \\ 
   & \downarrow  &  & \downarrow &  &  \downarrow  &  &   \\
   & 0  &  &  0  & & 0  & &   
\end{array}   \]

\[   \begin{array}{rcccccccl}
    &    &  &  0  &  &  0  &  &      \\
     &  &  & \downarrow &  &  \downarrow  &  &        \\
 &  0 & \rightarrow & 40 & \rightarrow & 40 & \rightarrow 0  \\
  & \downarrow &  & \downarrow & & \downarrow & &  \\
0 \rightarrow &  10 & \rightarrow  &  60 &  \rightarrow  &  50 &  \rightarrow 0   \\
    &  \downarrow &  & \downarrow &  &  \downarrow  &  &     \\
0 \rightarrow &  10 & \rightarrow  & 20 & \rightarrow  &  10  &  \rightarrow 0  \\ 
   &  &  & \downarrow &  &  \downarrow  &  &    \\
   &  &  &  0  & & 0  & &   
\end{array}   \]

\[   \begin{array}{rcccccccl}
    &    &  &  0  &  &  0  &  &   &   \\
     &   &  & \downarrow &  &  \downarrow  &  &   &     \\
     &  0  & \rightarrow & S_3 T^*\otimes T & \stackrel{{\sigma}_2(\Phi)}{\longrightarrow} &  S_2T^*\otimes S_2T^* & \rightarrow &   h_2 &  \rightarrow 0 \\
&  \downarrow &  & \downarrow &  &  \downarrow  &  &  \downarrow    &     \\
0 \rightarrow &  R_3& \rightarrow  & J_3(T) & \stackrel{{\rho}_2(\Phi)}{\longrightarrow}  &  J_2(S_2T^*) &  \rightarrow &  Q_2 &  \rightarrow 0  \\
    &  \downarrow &  & \downarrow &  &  \downarrow  &  & \downarrow   &     \\
0 \rightarrow &  R_2 & \rightarrow  & J_2(T) & \stackrel{{\rho}_1(\Phi)}{\longrightarrow}  &  J_1(S_2T^*) &  \rightarrow &  0 &  \\ 
   &  &  & \downarrow &  &  \downarrow  &  &  & \\
   &  &  &  0  && 0  &&   &
\end{array}   \]

\[   \begin{array}{rcccccccl}
    &    &  &  0  &  &  0  &  &   &   \\
     &  &  & \downarrow &  &  \downarrow  &  &   &     \\
 &  0 & \rightarrow & 80 & \rightarrow & 100& \rightarrow &  20 &  \rightarrow 0 \\
&  \downarrow &  & \downarrow &  &  \downarrow  &  &  \downarrow    &     \\
0 \rightarrow &  4 & \rightarrow  &  140 &  \rightarrow  &  150 &  \rightarrow &  14  &  \rightarrow 0  \\
    &  \downarrow &  & \downarrow &  &  \downarrow  &  & \downarrow   &     \\
0 \rightarrow &  10 & \rightarrow  & 60 & \rightarrow  &  50  &  \rightarrow &  0 &  \\ 
   &  &  & \downarrow &  &  \downarrow  &  &   & \\
   &  &  &  0  && 0  &&   &
\end{array}   \]

\[   \begin{array}{rcccccccl}
    &    &  &  0  &  &  0  &  &   &   \\
     &   &  & \downarrow &  &  \downarrow  &  &   &     \\
     &  0  & \rightarrow & S_4 T^*\otimes T & \stackrel{{\sigma}_3(\Phi)}{\longrightarrow} &  S_3T^*\otimes S_2T^* & \rightarrow &   h_3 &  \rightarrow 0 \\
&  \downarrow &  & \downarrow &  &  \downarrow  &  &  \downarrow    &     \\
0 \rightarrow &  R_4 & \rightarrow  & J_4(T) & \stackrel{{\rho}_3(\Phi)}{\longrightarrow}  &  J_3(S_2T^*) &  \rightarrow &  Q_3  &  \rightarrow 0  \\
    &  \downarrow &  & \downarrow &  &  \downarrow  &  & \downarrow   &     \\
0 \rightarrow &  R_3 & \rightarrow  & J_3(T) & \stackrel{{\rho}_2(\Phi)}{\longrightarrow}  &  J_2(S_2T^*) &  \rightarrow &  Q_2 & \rightarrow 0 \\ 
   &  &  & \downarrow &  &  \downarrow  &  & \downarrow  & \\
   &  &  &  0  && 0  && 0  &
\end{array}   \]

\[   \begin{array}{rcccccccl}
    &    &  &  0  &  &  0  &  &   &   \\
     &  &  & \downarrow &  &  \downarrow  &  &   &     \\
 &  0 & \rightarrow & 140 & \rightarrow & 200& \rightarrow &  60 &  \rightarrow 0 \\
&  \downarrow &  & \downarrow &  &  \downarrow  &  &  \downarrow    &     \\
0 \rightarrow &  2 & \rightarrow  &  280 &  \rightarrow  &  350 &  \rightarrow &  72  &  \rightarrow 0  \\
    &  \downarrow &  & \downarrow &  &  \downarrow  &  & \downarrow   &     \\
0 \rightarrow &  4  & \rightarrow  & 140. & \rightarrow  &  150  &  \rightarrow &  14 & \rightarrow 0 \\ 
   &  &  & \downarrow &  &  \downarrow  &  & \downarrow  & \\
   &  &  &  0  && 0  && 0  &
\end{array}   \]

Applying the Spencer $\delta$-cohomology to the top sequence, we get the commutative diagram in which the two central columns are exact: \\
\[  \begin{array}{rcccccccl}
   &  &  & 0  & &  0  &  & 0  &    \\
   &  &  &  \downarrow & & \downarrow & & \downarrow &   \\
   & 0 & \rightarrow &S_4T^*\otimes T& \rightarrow &S_3T^ *\otimes F_0 &\rightarrow & h_3  &  \rightarrow 0 \\
   &  &  &  \downarrow & & \downarrow & \searrow & \downarrow &   \\
   & 0& \rightarrow &T^*\otimes S_3T^*\otimes T& \rightarrow &T^*\otimes S_2T^ *\otimes F_0 &\rightarrow & \fbox{$ T^*\otimes h_2 $ }& \rightarrow  0  \\
   &   &  &  \downarrow & & \downarrow & & &  \\
   & 0  & \rightarrow &{\wedge}^2T^*\otimes S_2T^*\otimes T& \rightarrow &{\wedge}^2T^*\otimes T^ *\otimes F_0 &\rightarrow  &  0   \\
  &   &  &  \downarrow & & \downarrow & &  & \\
0 \rightarrow & \fbox{$ {\wedge}^3 T^*\otimes g_1 $ } & \rightarrow &{\wedge}^3T^*\otimes T^*\otimes T & \rightarrow &{\wedge}^3T^*\otimes F_0 &
    \rightarrow &   0  & \\
  & \downarrow  &  &  \downarrow & & \downarrow & &  &\\
0 \rightarrow &{\wedge}^4 T^*\otimes T & = &{\wedge}^4T^*\otimes  T& \rightarrow & 0 & &  \\
 & \downarrow  &  &  \downarrow & &  & & &  \\
 & 0  &  & 0  & &    &  &   &    
\end{array}  \]   \\  

\noindent
Using a (delicate) snake chase, we obtain the short exact sequence:  \\
\[  0 \longrightarrow h_3 \longrightarrow T^* \otimes h_2 \longrightarrow {\wedge}^3 T^* \otimes g_1 \longrightarrow {\wedge}^4 T^* \otimes T \longrightarrow 0, \hspace{5mm} 0 \rightarrow 60 \rightarrow 80 \rightarrow 24 \rightarrow 4 \rightarrow 0  \]
 and the long exact connecting diagram:  
\[ \begin{array}{rcccccccccl}
&  & 0 &  & 0 &  &  &  &  &  &   \\
&  &  \downarrow  &  & \downarrow &  &  &  &  &  &   \\
0 &\longrightarrow & S_4T^* \otimes T& \longrightarrow &S_3T^* \otimes F_0& \longrightarrow &T^* \otimes h_2 & \longrightarrow &H^3(g_1) &\longrightarrow &0 \\
  &   & \downarrow  &  &  \parallel & &  \downarrow  &  & \downarrow &  &  \\
 0 & \longrightarrow  & {\rho}_1(g'_2) & \longrightarrow  &S_3T^* \otimes F_0 & \longrightarrow &T^* \otimes Q_2& \longrightarrow & Q'_1 & \longrightarrow & 0 \\
   &  &  &  & \downarrow  &  & \downarrow  &  & \downarrow &  &  \\
    &  &  &  & 0 &  & 0 &  & 0 &  &
\end{array}  \]
allowing to compare the geometry of the Kerr metric (lower sequence) to the classical geometry of the Minkowski metric (upper sequence). In the later case we recall that the Bianchi indentities provide a section $ B \in H^3(g_1)$. Moreover, we have $dim(Q'_1) \leq dim(H^3(g_1))= 20$ if we define $Q'_1$ by the following commutative and exact diagram obtained by cutting the the fundamental diagram and going one step further on in the sequences in order to replace the $R$ and $Q$ by $R'$ and $Q'$:
\[  \begin{array}{rccccccccl}
    &  0  &  &  0  &  &  0  &  & 0  &   \\
     & \downarrow   &  & \downarrow &  &  \downarrow  &  & \downarrow  &     \\
 0 \rightarrow  & {\rho}_1(g'_2)  & \rightarrow & S_3T^*\otimes F_0 & \stackrel{{\sigma}_1(\Psi)}{ \longrightarrow }&  T^*\otimes Q_2 & \rightarrow &   h'_1 & \rightarrow 0 \\
&  \downarrow &  & \downarrow &  &  \downarrow  & \searrow &  \downarrow    &     \\
0 \rightarrow & {\rho}_1(R'_2) & \rightarrow  & J_3(F_0) & \stackrel{{\rho}_1(\Psi)}{\longrightarrow } &  J_1(Q_2) &  \rightarrow &  Q'_1 & \rightarrow 0   \\
    &  \downarrow &  & \downarrow &  &  \downarrow  &  & \downarrow   &     \\
0 \rightarrow &  R'_2 & \rightarrow  & J_2(F_0) & \stackrel{\Psi}{\longrightarrow}  &  Q_2 &  \rightarrow & 0 &   \\
   & \downarrow &  & \downarrow &  &  \downarrow  &  &   & \\
   & 0 &  &  0  &  & 0  &  &  &
\end{array}   \]
with dimensions:  \\
\[  \begin{array}{rccccccccl}
    &  0  &  &  0  &  &  0  &  & 0  &   \\
     & \downarrow   &  & \downarrow &  &  \downarrow  &  & \downarrow  &     \\
 0 \rightarrow  &  144 + x  & \rightarrow & 200 & \longrightarrow &  56 & \rightarrow &  x  & \rightarrow 0 \\
&  \downarrow &  & \downarrow &  &  \downarrow  & \searrow &  \parallel    &     \\
0 \rightarrow & 280 + x & \rightarrow  & 350 & \longrightarrow  &  70  &  \rightarrow &  x & \rightarrow 0   \\
    &  \downarrow &  & \downarrow &  &  \downarrow  &  & \downarrow   &     \\
0 \rightarrow &  136& \rightarrow  & 150 & \longrightarrow  &  14 &  \rightarrow & 0 &   \\
   & \downarrow &  & \downarrow &  &  \downarrow  &  &   & \\
   & 0 &  &  0  &  & 0  &  &  &
\end{array}   \]
Accordingly, the number $14$ of second order CC plus the number of differentially independent third order CC obtained by one prolongation of these second 
order CC is equal to $70 - x$. However, if we denote by $y$ the number of new generating CC of third order that are not differential consequences of the CC of second order already found, we must have $72 - y = 70 - x$ and thus $y = x + 2$.  \\
As we have $4$ divergence identities for the Einstein operator, we obtain therefore $ y = 4 + 2 = 6$ contrary to [1-4]. We have provided these new third order CC as $6$ components of the Spencer operator.  \\

Similarly, in the case of the S-metric, we have $ R^{(3)}_1 \subset R^{(2)}_1 \subset R^{(1)}_1 = R_1 \subset J_1(T)$ with respective dimensions $4 < 5 <10 = 10 < 20$ and we obtain:  \\
\[  Q_1 = 0, dim(Q_2) = 15, dim(Q_3) = 74 \Rightarrow 74 - y = 75 - x \Rightarrow  y = x - 1 = 3 \]
\[   \fbox { $d_1 {\xi}^1 - {\xi}^1_1=0, \,\,\, d_2 {\xi}^1 - {\xi}^1_2 = 0, \,\,\, d_3 {\xi}^1 - {\xi}^1_3 = 0  $ }    \]

In the case of the M-metric (See [7, 14, 15] for more details), we have $ R^{(3)}_1 = R^{(2)}_1 = R^{(1)}_1 = R_1 \subset J_1(T)$ with respective dimensions $ 10 = 10 = 10 < 20$ and we obtain:   \\
\[    Q_1 = 0, \,\,\, dim(Q_2) = 20, \,\,\, dim(Q_3) =  80, dim(Q'_1) =20 \rightarrow  x=20, y=0 \rightarrow  80 - y = 100 - x \]
in a coherent way with the number of components of the Riemann tensor ($20$) and the number of Bianchi identities ($20$) when $n=4$. The corresponding Killing system is FI if and only if the following Vessiot structure equations ${\rho}^k_{l,ij} = c ({\delta}^k_i {\omega}_{lj} - {\delta}^k_j {\omega}_{li})$ with the only constant $c$ are satisfied, that is the well known constant Riemannian curvature condition.  \\  \\

\noindent
{\bf 4) CONCLUSION} \\

\noindent
E. Vessiot discovered the so-called {\it Vessiot structure equations} as early as in 1903 and, only a few years later, E. Cartan discovered the so-called {\it Maurer-Cartan structure equations}. Both are depending on a certain number of constants like the single {\it geometric structure constant} of the constant Riemannian curvature for the first and the many {\it algebraic structure constants} of Lie algebra for the second. However, Cartan and followers never acknowledged the existence of another approach which is therefore still totally ignored today, in particular by physicists ([22-24]). Now, it is well known that the structure constants of a Lie algebra play a fundamental part in the Chevalley-Eilenberg cohomology of Lie algebras and their deformation theory ([14]). It was thus a challenge to associate the Vessiot structure constants with other homological properties related to systems of Lie equations, namely the extension modules determined by Lie operators. As a striking consequence, such a possibility opens a new way to understand and revisit the various contradictory works done during the last fifty years or so by different groups of researchers, using respectively Cartan, Gr\"{o}bner or Janet bases while looking for a modern interpretation of the work done by C. Lanczos from 1938 to 1962. However, the reader must not forget that the Weyl tensor was not known by Lanczos, even as late as in 1967, and that it was not possible to discover any solution of the parametrization problem by potentials through double duality before $1990/1995$, that is too late for the many people already engaged in this type of research ([20]). As for the study of the Killing operator, the intrinsic understanding of such a technical problem that we achieved has strictly nothing to do with GR. Indeed, thanks to a difficult theorem of homological algebra, the intrinsic properties of the resolution of a differential module $M$ only depend on $M$ itself but not on the resolution. Roughly, {\it the only important thing is the group, not the metric}. In actual practice, the resolutions {\it by themselves} no longer depend on the respective Kerr, Schwarzschild or Minkowski parameters. The same  comment can be done on the link existing between electromagnetism and the conformal group ([15, 21, 23]). We finally hope that this paper will open a new domain for applying computer algebra and we are offering a collection of useful but tricky test examples.  \\   \\

\noindent
{\bf REFERENCES} \\

\noindent
[1] Aksteiner, S., Andersson L., Backdahl, T., Khavkine, I., Whiting, B.: Compatibility Complex for Black Hole Spacetimes, Commun. Math. Phys. 384 (2021) 1585-1614.  \\
https://doi.org/10.1007/s00220-021-04078-y      (arXiv:1910.08756) .  \\
\noindent
[2] Aksteiner, S., Backdahl, T.: All Local Gauge Invariants for Perturbations of the Kerr Spacetime, Physical Review Letters 121, 051104 (2018), 
(arXiv:1803.05341) . \\
\noindent
[3] Aksteiner, S., Backdahl, T: New Identities for Linearized Gravity on the Kerr Spacetime, Phys. Rev. D 99, 044043 (2019), (arXiv:1601.06084) . \\
\noindent
[4] Andersson L., H\"{a}fner D., Whiting B.: Mode Analysis for the linearized Einstein equations on the Kerr Metric: The Large $ \mathfrak{a}$ Case, 2022 (arXiv:2207.12952).  \\
\noindent
[5] Goldschmidt, H.: Prolongations of Linear Partial Differential Equations: I Inhomogeneous equations, Ann. Scient. Ec. Norm. Sup., 4, 1 (1968) 617-625.  \\
\noindent
[6] Macaulay, F.S.: The Algebraic Theory of Modular Systems, Cambridge Tract 19, Cambridge University Press, 1916.  \\
\noindent
[7] Pommaret, J.-F.: Systems of Partial Differential Equations and Lie Pseudogroups, Gordon and Breach, New York, 1978; Russian translation: MIR, Moscow, 1983. \\
\noindent
[8] Pommaret, J.-F.: Differential Galois Theory, Gordon and Breach, New York, 1983. \\
\noindent
[9] Pommaret, J.-F.: Partial Differential Equations and Group Theory, Kluwer, Dordrecht, 1994.\\
https://doi.org/10.1007/978-94-017-2539-2    \\
\noindent
[10] Pommaret, J.-F.: Partial Differential Control Theory, Kluwer, 2001  (Zbl 1079.93001).  \\
\noindent
[11] Pommaret, J.-F.: Algebraic Analysis of Control Systems Defined by Partial Differential Equations, in "Advanced Topics in Control Systems Theory", Springer, Lecture Notes in Control and Information Sciences 311, 2005, Chapter 5, pp. 155-223.\\
\noindent
[12] Pommaret, J.-F.: The Mathematical Foundations of General Relativity Revisited, Journal of Modern Physics, 4 (2013) 223-239. \\
 https://doi.org/10.4236/jmp.2013.48A022   \\
 \noindent
[13] Pommaret, J.-F.: Relative Parametrization of Linear Multidimensional Systems, Multidim. Syst. Sign. Process., 26 (2015) 405-437.  \\
https://doi.org/10.1007/s11045-013-0265-0   \\
\noindent
[14] Pommaret, J.-F.: Deformation Theory of Algebraic and Geometric Structures, Lambert Academic Publisher (LAP), Saarbrucken, Germany, 2016 (arXiv:1207.1964).  \\
\noindent
[15] Pommaret, J.-F.: New Mathematical Methods for Physics, Mathematical Physics Books, Nova Science Publishers, New York, 2018, 150 pp. \\
\noindent
[16] Pommaret, J.-F.: Minkowski, Schwarzschild and Kerr Metrics Revisited, Journal of Modern Physics, 9 (2018) 1970-2007.  \\
https://doi.org/10.4236/jmp.2018.910125  (arXiv:1805.11958v2 ).  \\
\noindent
[17] Pommaret, J.-F.: The Mathematical Foundations of  Elasticity and Electromagnetism Revisited, Journal of Modern Physics, 10 (2019) 1566-1595, \\
https://doi.org/10.4236/jmp.2019.1013104  (arXiv:1802.02430) .  \\
\noindent       
[18] Pommaret, J.-F.: A Mathematical Comparison of the Schwarzschild and Kerr Metrics, Journal of Modern Physics, Journal of Modern Physics, 11 (2020) 1672-1710.  \\
https://doi.org/10.4236/jmp.2020.1110104   (arXiv:2010.07001).  \\
\noindent
[19] Pommaret, J.-F.: Minimum Parametrization of the Cauchy Stress Operator, Journal of Modern Physics, 12 (2021) 453-482.  \\
https://doi.org/10.4236/jmp.2021.124032  (arXiv:2101.03959).  \\
\noindent
[20] Pommaret, J.-F.: Homological Solution of the Lanczos Problems in Arbitrary Dimensions, Journal of Modern Physics, 12 (2021) 829-858. \\
https://doi.org/10.4236/jmp.2021.126053  (arXiv:1803.09610). \\
\noindent
[21] Pommaret, J.-F.: The Conformal Group Revisited, Journal of Modern Physics, 12 (2021) 1822-1842.  \\
https://doi.org/10.4236/jmp.2021.1213106  (arXiv:2006.03449 ).  \\
\noindent
[22] Pommaret, J.-F.: Minimum Resolution of the Minkowski, Schwarzschild and Kerr Differential Modules, Journal of Modern Physics, 13 (2022) 620-670.  \\
https://doi.org/10.4236/jmp.2022.134036 (arXiv:2203.11694).  \\
\noindent
[23] Pommaret, J.-F.: Nonlinear Conformal Electromagnetism, Journal of Modern Physics, 13 (2022) 442-494.  \\
https://doi.org/10.4236/jmp.2022.134031 (arXiv:2007.01710). \\
\noindent
[24] Pommaret, J.-F.: How Many Structure Constants do Exist in Riemannian Geometry ?.  \\
https://doi.org/10.1007/s11786-022-00546-3 (arXiv:2105.15126).  \\
\noindent
[25] Rotman, J.J.: An Introduction to Homological Algebra, (Pure and Applied Mathematics), Academic Press, 1979.  \\
\noindent
[26]  Spencer, D.C.: Overdetermined Systems of Partial Differential Equations, Bull. Amer. Math. Soc., 75 (1965) 1-114.\\
\noindent
[27] Vessiot, E.: Th\'{e}orie des Groupes Continus, Ann. Sc. Ecole Normale Sup., 20 (1903) 411-451 (Can be found in Numdam).  \\

\end{document}